\documentclass[useAMS,usenatbib]{mn2e}
\usepackage{graphicx,fleqn,times,color}
\usepackage{lscape}
\usepackage[T1]{fontenc}
\usepackage{aecompl}

\newcommand{\liaedit}{\color{black}}

\def\today{\ifcase\month\or
 January\or February\or March\or April\or May\or June\or
 July\or August\or September\or October\or November\or
 December\fi\space\number\day, \number\year}

\def\todmy{\number\day\space\ifcase\month\or
 January\or February\or March\or April\or May\or June\or
 July\or August\or September\or October\or November\or
 December\fi\space\number\year}

\newcommand{\tal}{\it et al. \rm}

\title[The barlens component in barred galaxies]{On the nature of the
  barlens component in barred galaxies: what do boxy/peanut bulges
  look like when viewed face-on?}

\author[E.~Athanassoula \tal] 
{
  E.~Athanassoula$^{1}$\thanks{E-mail:lia@lam.fr}, E. Laurikainen$^{2}$,
  H. Salo$^{2}$, A. Bosma$^{1}$\\ 
$^{1}$Aix Marseille Universit\'e, CNRS, LAM (Laboratoire
d'Astrophysique de Marseille),       
UMR 7326, 13388 Marseille 13, France\\
$^{2}$Department of Physics/Astronomy Division, University of Oulu,
FIN-90014, Finland\\
$^{3}$Finnish Centre of Astronomy with ESO (FINCA), University of Turku,
Vislntie 20, FI-21500 Pikki, Finland\\
}

\begin{document}

\date{Accepted . Received -}

\pagerange{\pageref{firstpage}--\pageref{lastpage}} \pubyear{2013}

\maketitle

\label{firstpage} 

\begin{abstract}

Barred galaxies have interesting morphological features whose presence
and properties set constraints on galactic evolution. Here we
examine barlenses, i.e. lens-like components whose extent along the bar
major axis is shorter than that of the bar and whose outline is
oval or circular. We identify and analyse barlenses in
$N$-body plus SPH simulations,
compare them extensively  with those from the NIRS0S (Near-IR S0
galaxy survey) and the S$^4$G samples (Spitzer Survey of Stellar
Structure in Galaxies) and find very good agreement. 
We observe barlenses in our simulations from different viewing
angles. This reveals that {\it barlenses are the vertically thick
part of the bar seen face-on, i.e. a barlens seen edge-on is a
boxy/peanut/X bulge}.
In morphological studies, and in the absence of kinematics or
photometry, a barlens, or part of it, may be mistaken
for a classical bulge. Thus the true importance of classical bulges,
both in numbers and mass, is smaller than currently assumed, which has
implications for galaxy formation studies. Finally, using the 
shape of the isodensity curves, we propose a
rule of thumb for measuring the barlens extent along the bar major
axis of moderately inclined galaxies, thus providing an estimate of 
which part of the bar is thicker. 

\end{abstract}

\begin{keywords}
galaxies: evolution -- galaxies: kinematics and
dynamics -- galaxies: spiral -- galaxies: structure 
\end{keywords}

\section{Introduction}
\label{sec:introduction}

To a zeroth order approximation, disc galaxies can be considered as
consisting of two simple components: an axisymmetric disc and a dark
matter halo. In the majority of cases, however, it is necessary 
to consider also at least a classical bulge and/or a bar, two components whose 
contribution to the total mass and light can be very significant and
which can play an important role in the dynamical evolution of the galaxy.
A more realistic picture will necessitate yet more components,
such as a stellar halo, a thick disc, spiral arms, rings and lenses. 

None of these components is simple. To take the bar as an
example, its description as an ellipsoid is in most cases
over-simplified, because the bar can include ansae \citep{Sandage.61,
  Laurikainen.SBK.09, Martinez.VKB.07},
can have a rectangular-like outline \citep{Athanassoula.MWPPLB.90,
  Gadotti.09, Gadotti.11} and, 
when seen edge-on, can have what is referred to as a boxy/peanut/X
bulge (hereafter the B/P/X bulge),
i.e. an inner part which is vertically thicker than the outer part
and has the shape of a box, a peanut, or an `X'. The formation and
evolution of this feature have been witnessed in a number of numerical
simulations (see \citealt{Athanassoula.15} for a review, as well as  
\citealt{Combes.Sanders.81}; \citealt{Combes.DFP.90};
\citealt{Pfenniger.Friedli.91}; \citealt{Raha.SJK.91}; 
 \citealt{Berentzen.HSF.98, Athanassoula.Misiriotis.02};
\citealt{Athanassoula.03, Debattista.CMM.04,Athanassoula.05}; \citealt{
   Martinez.VSH.06, Berentzen.SMVH.07}; etc.). 

Simulations of disc galaxies have given us a lot of information on the
formation and evolution of bars, the angular momentum exchanged in
barred galaxies and the secular evolution that ensues \citep[see][for
  a review, and references therein]{Athanassoula.13}. However,
in order to safely apply the results of these simulations to barred
galaxies and their observations, it is necessary to first make sure
that they reproduce well the observed bar properties, be they
morphological, photometrical, kinematical or, whenever the
simulations allow it, chemical. Morphological comparisons are
particularly important, since bar morphological properties have been
extensively studied (see \citealt{Buta.13a} and \citeyear{Buta.13b}
for reviews and references therein; see also
e.g. \citealt{Kormendy.79}, \citealt{Elmegreen.Elmegreen.85},
\citealt{Erwin.Sparke.03}, \citealt{Kormendy.Kennicutt.04},
\citealt{Buta.LSBK.06} and 
\citealt{Laurikainen.SBK.09}). In contrast, not many
simulation studies have focused on morphology. Nevertheless,
considerable progress was made by \cite{Athanassoula.Misiriotis.02},
who reproduced and discussed ansae and who also found that the bar
outline in strongly barred simulations was more rectangular-like that
elliptical-like, in good agreement with observations
\citep{Athanassoula.MWPPLB.90, Gadotti.09, Gadotti.11}.

\begin{figure*}
  \includegraphics[scale=0.4]{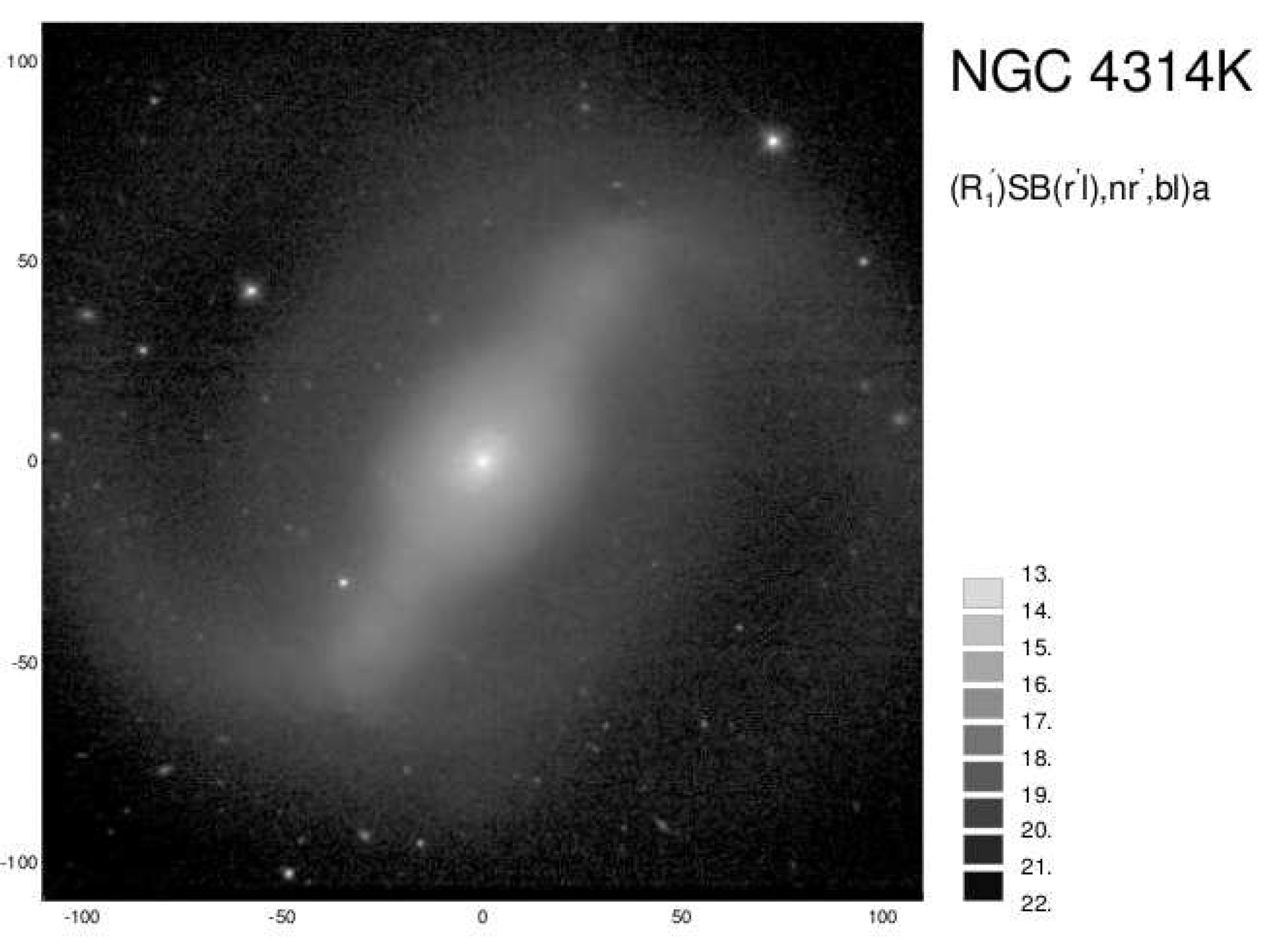}
  \caption{NIRS0S image of NGC 4314. The two components of the bar are
    clearly discernible. The inner component is shorter and fatter and
    the outer longer and more elongated (reproduced from
    \citealt{Laurikainen.SBK.11}, by permission of Oxford University
    press). The units on both axes are arcsec.} 
\label{fig:NGC4314}
\end{figure*}

In this paper, we will investigate a morphological feature of bars
which has only lately been outlined, namely the barlens (often
referred to as bl for short). \liaedit{These are lens-like components
  found in 
the central parts of barred galaxies and first unambiguously
identified, named and discussed by \cite{Laurikainen.SBK.11} in 
a study of the NIRS0S survey (Near-IR S0 galaxy Survey), where many
images of galaxies with barlenses can be seen (see also
\citealt{Laurikainen.Salo.15} for an observational review). 
Viewed face-on, their outline is oval to circular-like, in the direction 
along the bar major axis they are shorter than the
thin bar component \citep{Laurikainen.SABBJ.13} and perpendicular to
it they are more extended. They are distinct from nuclear lenses
because of  their much larger sizes and 
also from standard lenses \citep{Kormendy.79, Kormendy.Kennicutt.04}
because they are shorter than bars.} 
Further images of galaxies with barlenses can, in
  retrospect, be identified in e.g.
{\it The Hubble Atlas of Galaxies} \citep{Sandage.61}, in 
fig. 2 of \cite{Buta.LSBK.06} and in figs. 8 and 12 of \cite{Gadotti.08}.
They can also be seen in numerous publications based on the NIRS0S survey
\citep{Laurikainen.SB.05, Laurikainen.SBKSB.06, Laurikainen.SBK.07,
  Laurikainen.SBK.09, Laurikainen.SBKC.10}. 
Here we display in Fig.~\ref{fig:NGC4314} the NIRS0S image of NGC 4314, a
characteristic 
barlens example (reproduced from \citealt{Laurikainen.SBK.11}). In
this and many other cases one can see that the 
bar can be described as a sum of two components, a short, fat,
lens-like component, and 
a thin and much longer component. It is the former of these
components that is the barlens. A statistical
study of the NIRS0S sample provided useful information on the frequency with
which barlenses can be found in disc galaxies and on their sizes
\citep{Laurikainen.SABBJ.13}. These observational studies, however, were
not followed by any theoretical or simulation-based study. We thus ignore
whether the barlens is an independent component, or a 
morphological feature of the bar, and know nothing on how it forms and
evolves. We will aim at answering some of these questions in this
paper.

For this, we will use a number of high resolution simulations of bar
formation and evolution in disc galaxies, and compare the
morphology of their bars with that of the bars
in the NIRS0S and the S$^4$G samples. The NIRS0S sample 
\citep{Laurikainen.SBK.11} consists of about 200 nearby early-type galaxies,
mainly S0s, observed in the $K_S$ band using 3-4 m class telescopes
in good seeing conditions (full width at half-maximum $\sim$1 arcsec). The
sample selection criteria include the type ($-3 \le T \le 1$),
total magnitude ($B_T \le 12.5$) and inclination ($i \le 65^{\circ}$). The 
S$^4$G survey 
\citep{Sheth.p.10} is an 
Exploration Science Legacy Program carried out on the {\it Spitzer}
post-cryogenic mission. It includes more than 2300 galaxies from the
nearby Universe ($ D < 40$ Mpc) observed with the Infrared Array Camera
(IRAC) at 3.6 and 4.5 $\mu$. Its spatial resolution is
approximately 2 arcsec. 

This paper is organized as follows. In Sect.~\ref{sec:bar-struct} we
summarize theoretical results relevant to bar structure and morphology  
and in Sect.~\ref{sec:simulations} we present the simulations. In
Sect.~\ref{sec:global} we 
compare the morphology and the radial surface density profiles of real
to that of simulated galaxies and in Sects.~\ref{sec:ellipsefits} and
\ref{sec:decompositions} compare the results of the ellipse fits and
decompositions. We discuss our results in
Sect.~\ref{sec:discussion} and conclude in Sect.~\ref{sec:conclusions}. 
Further observational support for our work is presented in
an accompanying observational paper \citep[][hereafter
  L+14]{Laurikainen.SABHE.14}.      

\section{The structure of bars}
\label{sec:bar-struct}

When bars form they are vertically thin, no thicker than the disc 
they form in. This configuration, however, is vertically not stable 
\citep{Binney.78} and, after a relatively short time, the inner part
of the bar thickens considerably and protrudes clearly out of the
galactic disc (for a review see \citealt{Athanassoula.15} and also
references therein and in Sect.~\ref{sec:introduction}).
It is important to stress that it is {\it not} the whole bar that
thickens, but only the inner parts, the outer parts remaining thin 
\citep{Athanassoula.05}. The bar can thus be considered as consisting
of two parts, an outer and an inner part, the former being thin both
in the equatorial plane and vertically to it, and the latter being
thick in both these directions. Let us now check how
theoretical work and observations compare with this picture.

This bar structure can be easily understood with the help of 3D orbital
structure \citep*{Pfenniger.84, Skokos.PA.02a, Skokos.PA.02b}. The
backbone of the bar is the $x_1$ family of orbits, which 
are periodic -- closing after one revolution and two radial
oscillations -- planar and elongated along the bar 
\citep{Contopoulos.Papayannopoulos.80, Athanassoula.BMP.83}. They are stable
over most of the family extent, and thus trap regular orbits around
them. This family 
has, nevertheless, a number of vertical instabilities, from which bifurcate
other families, often called $x_1v_1$, $x_1v_2$, $x_1v_3$, etc.
\citep{Pfenniger.84, Skokos.PA.02a, Skokos.PA.02b}. Their orbits
have shapes which, projected on the equatorial plane, are 
similar to those of the $x_1$ family orbits, but, contrary to those of
the $x_1$, they can have a considerable vertical 
extent. Some of these families have important stable parts, and can thus
trap regular orbits around them. These constitute the B/P/X bulge
and orbital structure suggests that its major axis length is shorter than
that of the bar \citep*{Patsis.SA.02}. This has been well confirmed
and established by simulations \citep[][and later]{Athanassoula.05} by
measuring the length of the B/P/X bulge in a side-on\footnote{In the
  side-on view the galaxy is 
observed edge-on and the line of sight is on the equatorial
plane along the minor axis of the bar.} view, and the length
of the bar in a face-on one and comparing the two lengths. Their
ratio, according to orbital structure theory, can take values within a
wide range, depending on which perpendicular family sets the
extent of the structure. 
The value of this ratio was considered for the case of M31 by
\cite{Athanassoula.Beaton.06} and will be further discussed here for a
number of $N$-body simulations. Let us now briefly
review the input from observations.

In our Galaxy the bar is seen edge-on and, due to the location of the
sun, is seen only after an integration
along lines of sight spanning an angular extent of several degrees. For these
reasons it is not possible to have a clear 
view of the bar morphology, so that the thin and the thick bar components
were initially considered as two independent bars, with, moreover,
two independent position
angles and were thus assumed to rotate at different pattern speeds
\citep{Hammersley.GMLC.00, Benjamin.P.05, Cabrera.LHGFLCGM.07,
  Lopez.CCLMHGGF.07, Cabrera.LGFGHLC.08,
  Churchwell.BMWBICRPWB.09}. More recent work, 
however, argued clearly that these 
two bars can simply be the thin and the thick part of a single bar
\citep{Athanassoula.06, Romero.GAAF.11, Martinez.VG.11}. 

Boxy/peanut bulges can of course be easily distinguished in edge-on or
near-edge-on external galaxies, since they clearly protrude out of the
disc plane. 
Using photometry, it is also possible to show that there 
is an associated thin component, embedded in the
disc. \cite{Lutticke.DP.00} and \cite{Bureau.AADBF.06} used infrared
images of samples of 60 and 30 edge-on galaxies, respectively, and
analysed the luminosity profile from narrow slits along or parallel
to, but offset from, the major axis of each galaxy. They could thus
verify the existence of the thin and the thick part of the bar and
measure their relative extents, as well as some further relevant
quantities. In several cases, however, this task can be rendered
very complex by the existence of other thin components in the
disc, such as lenses or rings \citep[e.g.][]{Wakamatsu.Hamabe.84,
  Kormendy.Kennicutt.04}.  
Unexpectedly, it is in inclined, but not necessarily edge-on, galaxies
that the existence of the two parts of the bar can be seen easiest.   
This was first noted for NGC 4442 by 
\cite{Bettoni.Galletta.94}, for NGC 7582 by \cite{Quillen.KFD.97} and for
M31 by \cite{Athanassoula.Beaton.06} and \cite{Beaton.MGSCGPAB.07},
galaxies which have 
an inclination of 72$^\circ$, 65$^\circ$ and 77$^\circ$,
respectively. A much larger sample of 78 galaxies, including
intermediate inclination angles, was analysed by 
\cite{Erwin.Debattista.13}. 

$N$-body simulations are of course particularly well suited for 
studying components with such a complex shape as bars, because they
can be viewed from any desired angle. Thus, the extent of the thin part
of the bar can be measured from the face-on view, while that of the
thick part from the side-on one. The snapshot can then be 
viewed from intermediate inclination angles
in order to educate the eye to distinguish the existence and
the extent of the thin and thick parts of the bar. This was first
performed on simulations by \cite{Athanassoula.Beaton.06},
in order to obtain the necessary expertise to distinguish the bar in
M31 and get information on its properties, and later by
\cite{Erwin.Debattista.13}, who
 extended this work to intermediate inclination angles. 

But what about the
near-face-on, or moderately inclined galaxies? Answering this 
question is a primary goal of our paper. 

\section{Simulations}
\label{sec:simulations}

For comparisons with observations, we will use two groups of
simulations initially made to study the effect of gas and of halo
properties on bar formation and evolution. The first one includes
simulations with gas and with 
spherical or triaxial haloes, each one of which will be denoted by 6
characters. The first three are `gtr' ({\bf g} for {\bf g}as and {\bf
  tr} for {\bf 
  tr}iaxial), and are common to all simulations of the group. The
last three form a number which distinguishes the various simulations
of the group between them. These simulations were described in
detail in (\citealt*{Athanassoula.MR.13}, hereafter AMR13) so we will 
describe them here only briefly. The second group is designated by the
first three letters `gcs' ({\bf g} for {\bf g}as and {\bf cs} for {\bf
  c}u{\bf s}p) and has many points in common with the first
group (Athanassoula, in prep.). 
In their
initial conditions, all gtr and gcs simulations have a disc and a halo
component, the main difference between the two groups being that the
halo of the gtr has a small core, while that of the gcs a
cusp\footnote{In radial profiles with a core, the halo density levels off 
  as one reaches the centre, while in cuspy profiles it keeps
  increasing with decreasing radius.}. 

All simulations were run with the \textsc{gadget} code
\citep{Springel.YW.2001, Springel.Hernquist.02, Springel.Hernquist.03,
  Springel.05}, a softening of 50 pc for all components, and a
cell-opening criterion corresponding to an error tolerance for the force of
0.005\footnote{For more information on the algorithmic and numerical aspects
of \textsc{gadget} and the exact definition of its parameters see the
manual in http://www.mpa-garching.mpg.de/$\sim$volker/gadget  and  
the above mentioned references.}.
The \textsc{gadget} code offers the possibility of using several types of
particles for the various components of the galaxy. In the following we
will use four types, namely: \textsc{halo}, \textsc{disk},
\textsc{gas} and \textsc{stars}. The \textsc{disk} particles represent stars
already present in the initial conditions and their number 
remains constant throughout the simulation. But as the simulation
evolves, new stars (\textsc{stars}) form from gas (\textsc{gas}), so that
the number of \textsc{gas} particles decreases, while that of
\textsc{stars} increases.  
At any time, the \textsc{disk} particles and the   
\textsc{star} particles together represent the stars of the galaxy,
the \textsc{disk} component describing older stars than the 
\textsc{stars} component. But all the \textsc{stars} do not represent
young stars. Indeed, a fair fraction of them are formed during the
first two Gyr and thus have similar age and dynamics to that of 
\textsc{disk} particles. The main difference between the two
components is that all the \textsc{disk} particles represent the old
stellar population, while a fraction of the \textsc{stars} describe
the young stellar population. This fraction, however, is a strongly
decreasing function of time (see Fig. 2 of AMR13), because the
simulations do not include accretion. 

The \textsc{disk} has in all cases the same functional form for the
initial azimuthally 
averaged density distribution, namely

\begin{equation}
\rho_d (R, z) = \frac {M_d}{4 \pi h^2 z_0}~~exp (- R/h)~~sech^2 (\frac{z}{z_0}),
\end{equation}

\noindent
where $R$ is the cylindrical radius, $M_d$ is the disc mass, $h$ is the
disc radial scalelength and $z_0$ is the disc vertical scale
thickness.
It also has the same initial radial velocity dispersion profile, namely

\begin{equation}
\label{eq_svR}
\sigma_R(R) = 100 \cdot \exp\left(-R/3h\right) \; {\rm km \, s^{-1}} \, . 
\end{equation}

For all gtr runs the initial $\textsc{disk}$ component has a radial
scalelength of $h=3$~kpc, a scale  
height of $z_0=0.6$~kpc, and the mass of each individual particle  
is $m_{\textsc{disk}}= 2.5 \times 10^{5}~M_{\odot}$. 
The corresponding numbers for the gcs simulations are $h=4$~kpc, 
$z_0=0.6$~kpc, and $m_{\textsc{disk}}=6.25 \times 10^{4}~M_{\odot}$.

For the gas, we adopt the same radial profile and the same scalelength
as for the stellar disc. This is necessary in order to be able to make 
sequences of models where only the gas fraction changes, while the
rotation curve stays the same for all practical purposes. The gas scaleheight
is considerably smaller than that of the stars and its precise value is
set by the hydrostatic equilibrium achieved during the iterative
calculation of the initial conditions \citep{Rodionov.Athanassoula.11}.

For gtr simulations the 
haloes have been built so as to have, within the allowed accuracy, 
in all cases the same spherically averaged initial radial profile, namely
 
\begin{equation}
\rho_h (r) = \frac {M_h}{2\pi^{3/2}}~~ \frac{\alpha}{r_c} ~~\frac
{exp(-r^2/r_c^2)}{r^2+\gamma^2},
\end{equation}

\noindent
where $r$ is the radius, $M_h$ is the mass of the halo and $\gamma$
and $r_c$ are the halo core and cut-off radii, respectively. The
parameter $\alpha$ is a normalization constant defined by 

\begin{equation}
\alpha = [1 - \sqrt \pi~~exp (q^2)~~(1 -erf (q))]^{-1},
\end{equation}

\noindent
where $q=\gamma / r_c$ \citep[][]{Hernquist.93}. All simulations have
$\gamma=1.5$~kpc, $r_{c}=30$~kpc and $M_{h}=2.5 \times 10^{11}~M_{\odot}$.
The mass of each halo particle is equal to $m_{halo}= 2.5 \times 10^{5}~M_{\odot}$.
In this group of simulations, we consider three different initial halo
shapes: spherically 
symmetric (halo 1), mildly triaxial with initial equatorial and vertical axial
ratios 0.8 and 0.6 (halo 2), respectively, and strongly triaxial with 
initial axial ratios 0.6 and 0.4 (halo 3), respectively. As shown in AMR13,
these haloes evolve with time and their shapes become near spherical for halo2
and mildly triaxial for halo 3. 
 
For all gcs simulations, the dark halo density profile follows that of
a truncated NFW halo (\citealt*{Navarro.FW.96,Navarro.FW.97}) 
\begin{equation}
\rho_{\rm h}(r) =  \frac{C_{\rm h} \cdot T(r/r_t)}
{(r/r_h)(1+r/r_h)^2} \, ,
\label{eq_NFW}
\end{equation}

\noindent
where $T(r/r_t) = exp ( - r^2 / r_t^2) $,
$r_h$ is the halo scalelength, $C_h$ is a parameter defining
the mass of the halo, and $r_t$ is a characteristic radius for the
tapering. For all gcs models discussed here $r_h=11\;\rm kpc$ and
$r_t=50 \;\rm kpc$. For halo 4 models $C_h=0.0015$ and for
halo 5 ones $C_h=0.002$. In each of these simulations there are two million
particles in the halo. 

The initial conditions of both groups were built using the iterative
procedure \citep*{Rodionov.AS.09, Rodionov.Athanassoula.11}. The
corresponding rotation curves of gtr models 
are given in Fig. 1 of AMR13. The list of the runs analysed here, together 
with their initial gas fraction and halo type are listed in
Tables~\ref{tab:rellipsefits} and \ref{tab:rdecompositions}. To
measure the bar strength during the evolution, we use 
in all cases the maximum of the relative $m$=2 Fourier component, as
described in AMR13. 

The next step is to use snapshots of the stellar disc component
in order to create mock images in form of fits files, which will be
analysed in 
the same way as the observed fits images of galaxies.
Due to the relatively small differences between the density
distributions of the \textsc{disk} and \textsc{stars}, we used these
two components together in order to 
improve the signal-to-noise ratio by increasing the number of
particles. Furthermore, as already mentioned above, most of the stars
were born in the first few Gyr of the simulation, so that the
\textsc{disk} and \textsc{stars} together amount to a population
compatible with the light emitted by disc galaxies in the near- or
mid- IR \citep{Meidt.plus.12a, Meidt.plus.12b, Meidt.plus.14,
  Querejeta.plus.14}. 
The analysis consists mainly of ellipse fits and
decompositions, and the techniques and results are described in
Sect.~\ref{sec:ellipsefits} and \ref{sec:decompositions}. 

\section{Global comparisons: morphology and radial surface density profiles}
\label{sec:global}

\begin{figure*}
  \includegraphics[scale=0.8]{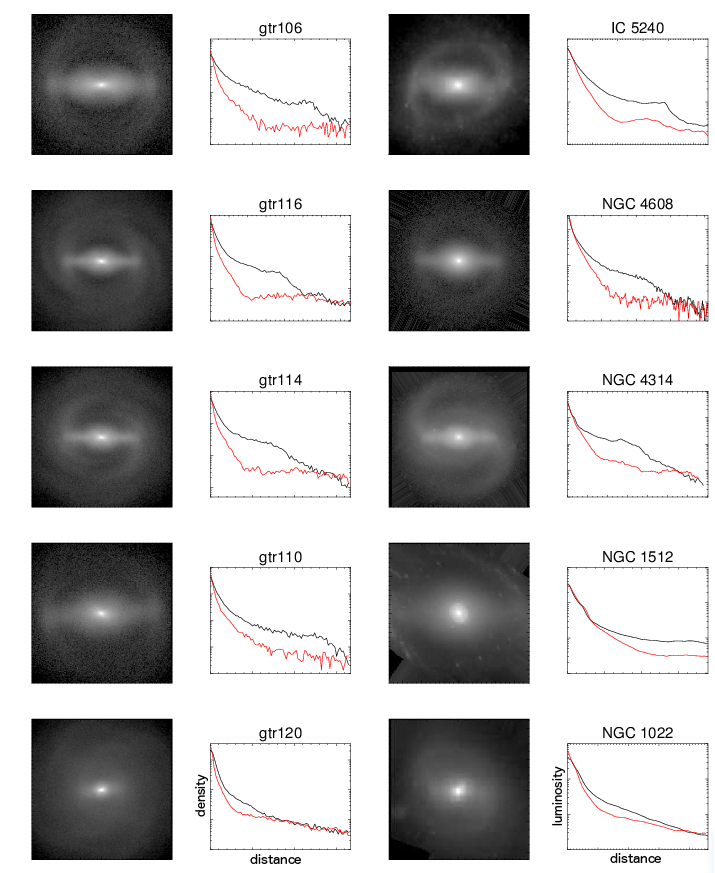}
  \caption{
Images and the corresponding surface brightness profiles shown along
  the bar major (black lines) and minor axis (red lines in the online
  version). The two left-hand columns
  show the simulation models and the two right ones galaxies with
  similar barlenses. The extent of the images has been chosen so as to
  display in all cases the same region of the galaxy, namely the bar
  region and, 
  whenever relevant, the inner ring. The galaxy images are
  either $K_s$-band images from the NIRS0S Atlas, or 3.6 $\mu$m images
  from the S$^4$G survey. Both the simulations and the galaxies are
  shown in face-on view, and rotated so that the bar
  appears in a horizontal position. The profiles cover the same
  regions as the corresponding images. The name of the galaxy, or of
  the simulation is given above the corresponding radial surface
  brightness panel. The units of distance, density and  
  luminosity are arbitrary (see text).
    }
\label{fig:comp-morpho}
\end{figure*}

A simple perusal of the face-on views of the gtr snapshots shown in
figs. 4 and 5 in AMR13 
shows clearly that, like in observations, simulated bars are not 
simple ellipsoidal objects, but have a more complex geometry. In 
the equatorial plane, the inner part of the bar is considerably thicker 
than the outer thin bar part and has an oval or near-circular shape. A 
more detailed visual inspection shows that, the morphology
of these central components is such that,  
had they been observed in galaxies, they would have been classified as
barlens components. In the rest of this paper we will
deepen and substantiate the comparison between barlenses in real
galaxies and their
counterparts in simulations, and, after establishing that the
latter should be called barlenses, we will examine in detail their
properties in order to get more information on the barlens structures. 
  
\begin{figure*}
  \includegraphics[scale=0.4]{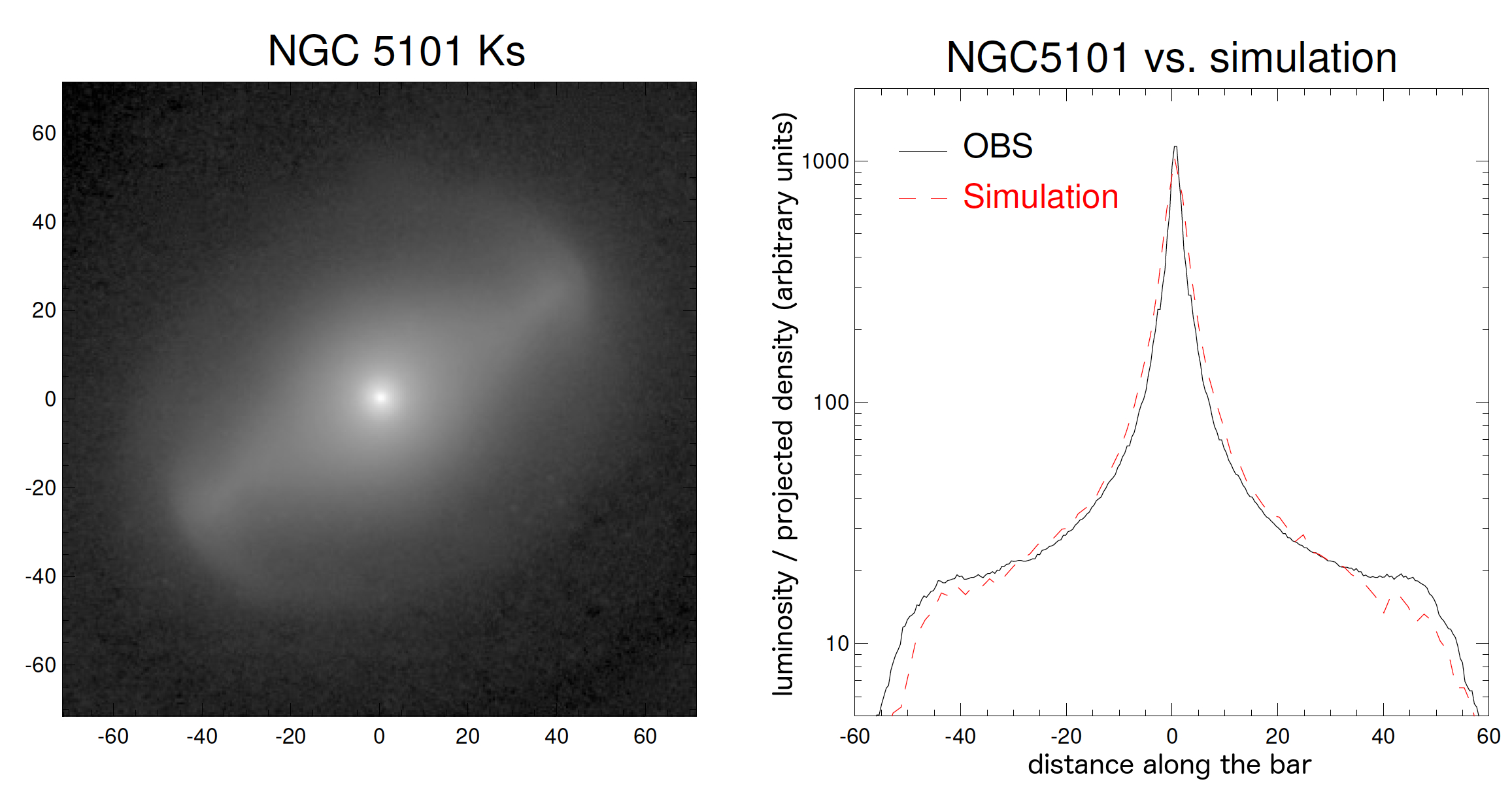}
  \caption{Left-hand panel: the bar region of NGC 5101 in the Ks band (from
    the NIRS0S sample). Right-hand panel: 
the surface brightness profiles of the simulation model gtr115 at $t$
= 6 Gyr (dashed line, red in the online version) and NGC 5101
(full black line), both shown along the bar
major axis. As in the previous figure, only the bar region is shown,
so that the sharp edge 
in the profile is a manifestation of the edge of the bar. The units
for the NGC 5101 figure, as well as for the corresponding distance
along the bar (black line) are in arcsec, while those of the
simulations, are arbitrary. The units for projected density of the simulations
as well as for the luminosity for the observations, are arbitrary (see
text). 
    }
\label{fig:NGC5101_gtr115_prof}
\end{figure*}

Let us start with comparisons including the morphology
and the radial density profiles (Figs.~\ref{fig:comp-morpho}, 
\ref{fig:NGC5101_gtr115_prof} and \ref{fig:comp-NGC5101_gtr115}).   
Fig.~\ref{fig:comp-morpho} compares the results from five simulated
galaxies (two left-hand columns) to those of five galaxies from the NIRS0S,
or from the S$^4$G surveys (two right-hand columns). 
We carried out this exercise for many more snapshots and galaxies, but we
will limit ourselves here to five for brevity.
It is important to note that the five simulations shown in
Figs.~\ref{fig:comp-morpho} were 
not run specifically so as to model the five chosen galaxies, they
were simply picked out by a cursory visual inspection of the images of
several snapshots of the gtr and gcs simulations
as reasonable matches to the five chosen
galaxies. Nevertheless, we get a good resemblance, both for the images and
the radial projected density profiles in the region we examine, i.e. a
rectangular region of linear extent somewhat larger than the bar
region and including, whenever relevant, the inner ring.

The upper row compares simulation gtr106, viewed at a time $t$=6 Gyr
after the beginning of the simulation, with an S$^4$G image of the galaxy
IC 5240. This comparison involves two of the longest and most elongated
barlenses and quite strong bars. As can be seen in the second panel
from the left, the bar projected surface
density profiles are quite flat, arguing for a strong bar (see
\citealt{Elmegreen.Elmegreen.85} and \citealt{Kim.plus.15} for the
observations and 
\citealt{Athanassoula.Misiriotis.02} for the simulations). Indeed, it
was found in AMR13 that gtr106 was one of the strongest bars in that
set of simulations (see in particular figs. 4, 5, 7 and 8 in that
paper and the
corresponding discussions). In this example the extent of the thin bar
outside the barlens component is very short, amongst the shortest in
our sample. In both the
observed and the simulated galaxy profiles this manifests itself by a
ledge of very short extent. The part of the profile that is within the
ledge rises inwards, but much less steeply than that of classical
bulges. This will be discussed further in Sect.~\ref{subsec:NGC936}. 
Note also that in both gtr106 and IC 5240,
there is an X-shape in the central regions, even though
the simulated galaxy is seen exactly face-on and IC 5240 quite far from
edge-on. Such a behaviour is discussed in detail in the accompanying
observational paper (L+14).

The second row from the top compares the image of the $t$=6 Gyr snapshot of gtr116
with the NIRS0S image of NGC 4608. In this case, the relevant extent of
the thin part of the bar --
relative to both the extent of the barlens and the size of the inner
ring -- is much longer than in the previous example.   
Note also how thin in the equatorial plane
the part of the bar outside the barlens is in the (near-)face-on views of
both the galaxy and the simulation snapshot. Further discussion on NGC
4608 is given in Sect.~\ref{subsubsec:decomp-comparisons}.

The third row compares the $t$=6 Gyr image of gtr114 with the S$^4$G
image of NGC 4314. Both have a fairly round barlens and a strong thin
bar component. In the fourth row we compare
gtr110 at $t$=6 Gyr with the S$^4$G image of NGC 1512. Here the barlens is
similar to that in the two previous models, but the thin bar is
considerably less pronounced, and more dispersed.

Finally, in the bottom row we compare the $t$=6 Gyr image of gtr120 to
the S$^4$G image of NGC 1022. This is the weakest bar in this comparison, 
and one of the weakest bars in the AMR13 simulations. This can be
witnessed also from the radial surface density profiles, which shows
clearly that the difference between the profiles along the directions
of the bar major and 
minor axes is less than in the previous cases and that the bar has
an exponential profile. The latter according to
\cite{Athanassoula.Misiriotis.02} is a sign of a bar weaker than 
those having a flat profile. In contrast, the barlens is still very
prominent.

The right-hand panel of Fig.~\ref{fig:NGC5101_gtr115_prof} compares the
radial projected 
surface density profiles along the direction of the major axis of the bar 
of gtr115 and NGC 5101. Their barlens regions look remarkably
similar, although the 
former was not made specifically so as to model the latter. There is
also a further point to note in this figure, which can also be seen in
the radial profiles of Fig.~\ref{fig:comp-morpho}. The end of the thin
bar component can be easily distinguished on these profiles because of
a clear jump, or step, of the radial profile at this radius. On the
contrary, the profile has no feature which could indicate the
existence of a barlens or its extent.
This is presumably because the thin part of
the bar extends inwards into the barlens region and thus the profile along
the bar major axis includes contributions from both components.
It will be discussed further in Sect.~\ref{sec:discussion}. 

\section{Ellipse fitting: Technique and results}
\label{sec:ellipsefits}

\subsection{Techniques}
\label{subsec:ell-technique}

We chose a fitting procedure which is identical to that used for the
NIRS0S galaxies, in order to allow comparisons.
Given the complex morphological structure of bars, any automatic
ellipse fitting would not give useful information concerning the
barlens component. For example, isophotes corresponding to the outer
regions of the barlens will also trace regions of the outer part of the
thin bar component and any automatic ellipse fit would be a compromise
between the shape of the 
barlens and that of the thin outer parts of the bar. Thus, such fits
would be strongly biased and useless for our purposes. We therefore
resorted to a 
fully interactive fitting procedure, which, although much more time
consuming, guarantees relevant fits. 

For the simulated galaxies, before starting
the measurements, the simulation units of the image were converted to
magnitudes by logarithmic transformation, using a magnitude zero-point
of 24 mag/arcsec$^2$. The subsequent procedure was followed
{\it identically} for observed and for simulated galaxies. It is very
similar to that of masking, where regions which are heavily influenced 
by other components are masked out. We first inspected all images to
determine the magnitude range in which the barlens is 
best visible, and this was systematically used in all further
measurements. We then visualized each individual galaxy separately 
and chose by eye points on the outermost edge of the barlens which are
not affected by the thin part of the bar. Thus these points avoid the
region around the direction of the bar major axis. An ellipse was
fitted to these points using the same software as for the NIRS0S
galaxies. Since the thin part of the bar is thin not only vertically
but also on the equatorial plane, the eliminated part is rather
restricted and the number of points and the area that is left after
elimination are well sufficient to allow an adequate fit, both for
observations and for simulations.

Our approach has two important advantages. First, 
the measurements are not affected by the narrow outer
part of the bar (thin bar component), which in
automatic ellipse fitting would erroneously make the barlens too
elongated. Secondly, it is identical to the one used for the
observations, thus best allowing comparisons.

In order to obtain optimum conditions for comparisons of simulations
and observations, as well as to minimize the chances of any humanly
introduced bias, we 
proceeded as follows. One of us (EA) chose the snapshots to be
analysed in view of the specific goals of our paper, and prepared the
fits files. Both the ellipse fits and the decompositions were made by
another member of the team (EL), who had also done this same analysis for
the NIRS0S sample galaxies. The full description of the runs and of the
corresponding snapshots was communicated to the analyser only after she
had completed the analysis. As a 
result, some of the analysis was done two or three times. For example all
snapshots in Table~\ref{tab:tellipsefits} were analysed together and we
did not simply carry over the results from
Table~\ref{tab:rellipsefits} for $t$=6 cases. In this way, some of the
snapshots were measured independently two, or three times. A
posteriori, we use 
this to set some estimates of the errors in the ellipse fits. We note
that for most cases the differences between the two independent
estimates of the same snapshots and quantities, are quite small,  
less or of the order of 10\%, showing that the 
measurements are repetitive, as they should. There is, however, one
case, namely gtr101 at $t$=6 Gyr, where the error reaches roughly
20\%. This is a very difficult case to measure and, as will be
discussed in the next section, we were also unable to make a good
decomposition. Furthermore, as will be discussed in
Sect.~\ref{sec:discussion}, the barlenses in the gas-less 
simulations are less realistic than those in the remaining runs, with
more elongated shapes than usually observed. It is thus not surprising
that the parameters of this one case obtained from the ellipse fits are not as robust
as those of the other models.

\begin{table}
   \caption[]{Results from our visually driven ellipse fits (see
     Sect.~\ref{subsec:ell-technique}) for the barlenses (bl) and the
     thin part of the bar (bar) in simulation results at time $t$=6
     Gyr. }
\begin{tabular}{ccccc}
\hline
\noalign{\smallskip}
\noalign{\smallskip}
  run & Halo profile & initial & b/a (bl) & r$_{bl}$/r$_{bar}$ \\
  &  &  gas [\%] &  &   \\
 \noalign{\smallskip}
\noalign{\smallskip}
\hline
 \noalign{\smallskip}
\noalign{\smallskip}
gtr101  & 1: core/spherical & 0  & 0.43 & 0.77    \\
gtr106  & 1: core/spherical & 20  & 0.65 & 0.57    \\    
gtr111  & 1: core/spherical & 50  & 0.95 & 0.37    \\   
gtr116  & 1: core/spherical & 75  & 0.93 & 0.45    \\   
gtr119  & 1: core/spherical & 100  & 0.74 & 0.56    \\  
 \noalign{\smallskip}
gtr102  & 2: core/mildly triaxial & 0  & 0.45 & -      \\
gtr109  & 2: core/mildly triaxial & 20  & 0.74 & 0.47   \\
gtr114  & 2: core/mildly triaxial & 50  & 0.88 & 0.46   \\ 
gtr117  & 2: core/mildly triaxial & 75  & 0.89 & 0.81   \\
gtr120  & 2: core/mildly triaxial & 100  & 0.94 & 0.51   \\
 \noalign{\smallskip}
gtr110 & 3: core/strongly triaxial & 20 & 0.68 & 0.49   \\   
gtr115 & 3: core/strongly triaxial & 50 & 0.93 & 0.52   \\    
gtr118 & 3: core/strongly triaxial & 75 & 0.88 & 0.53   \\    
gtr121 & 3: core/strongly triaxial & 100 & 0.88 & -      \\    
\noalign{\smallskip}
\noalign{\smallskip}
gcs006 & 4: cusp1/spherical & 0 & 0.64 & 0.94   \\
gcs001 & 4: cusp1/spherical & 20 & 0.89 & 0.46   \\
gcs002 & 4: cusp1/spherical & 40 & 0.96 & 0.44   \\
gcs003 & 4: cusp1/spherical & 60 & 0.90 & 0.52   \\
gcs004 & 4: cusp1/spherical & 80 & 0.95 &  -     \\
\noalign{\smallskip}
gcs009 & 5: cusp2/spherical & 60 & 0.74 & 0.56   \\
gcs010 & 5: cusp2/spherical & 80 & 0.91 & 0.46  \\
 \noalign{\smallskip}
\hline
\end{tabular}
\label{tab:rellipsefits}
\end{table}

\vfill
\eject

Bar lengths were measured in a similar manner by visually
marking the outer edges of the bar. When the simulation 
model has an inner ring there is some ambiguity, because the bar can be
considered to end where the ring starts (i.e. at the inner radius
of the ring ), or to continue all through
the ring, (i.e. to end at the outer radius of the ring), or to
end at 
some intermediate position. Which of these estimates is best is
debatable, and may even depend on the case at hand. Here we follow
the same definition as in the NIRS0S observations,
namely that the end of the bar is defined by the ridge-line of the
ring, in order to allow comparisons between simulations and observations, 
particularly since the bar length estimate will be used
further in the decompositions (see Sect.~\ref{sec:decompositions}).   


\begin{table}
   \caption[]{Results of the axial ratio and relative radial extent of
     the barlens (bl) component for three simulations, as obtained from
     our visually driven ellipse fits (see 
     Sect.~\ref{subsec:ell-technique}). 
}
\begin{tabular}{lccccccc}
\hline
\noalign{\smallskip}
\noalign{\smallskip}
Time [Gyr] -$>$       & 5 & 6 & 7 & 8 & 9 & 10 \\
\noalign{\smallskip}
\noalign{\smallskip}
\hline
\noalign{\smallskip}
\noalign{\smallskip}
{\it gtr101:} & & & & & & & \\
\noalign{\smallskip}
b/a(bl)       &0.59 &0.53 &0.50 &0.47 &0.44 &0.43 \\
r$_{bl}$/r$_{bar}$  &0.61 &0.78 &0.86 &0.84 & - & - \\
\noalign{\smallskip}
{\it gtr111:}       & & & & & & & \\
\noalign{\smallskip}
b/a(bl)      &0.93 & 0.95&0.53 &0.64 &0.79 & 0.83\\
r$_{bl}$/r$_{bar}$ &0.39 &0.34 &0.74 &0.69 &0.56 &0.57 \\
\noalign{\smallskip}
{\it gtr119:}       & & & & & & & \\
\noalign{\smallskip}
b/a(bl)      &0.72 &0.69 &0.79 &0.79 &0.72 & 0.75\\
r$_{bl}$/r$_{bar}$ & - & 0.61&0.53 &0.61 &0.64 &0.61 \\
\noalign{\smallskip}
\hline
 \noalign{\smallskip}
\end{tabular}
\label {tab:tellipsefits}
\end{table}

\subsection{Results}
\label{subsec:ell-results}

Table \ref{tab:rellipsefits} gives information on the ellipses fitted to
the face-on images of the simulation snapshots at time $t$ = 6 Gyr from the
beginning of the run. The first three columns give the simulation
name, its initial halo characteristics and the initial fraction of its
baryons that is in the form of gas,
respectively. The fourth and fifth columns give the axial ratio 
of the barlens component and its extent along the bar major axis
normalized by the bar length. The five packages in this Table
(separated by a blank horizontal line) correspond to the five
different halo radial 
profiles, halo 1 to halo 5, described in Sect.~\ref{sec:simulations}.
 Results from  different 
viewing angles will be discussed later (see Table~\ref{tab:view-decomp}
and Sect.~\ref{subsubsec:test-decomp}).

In this, as in all subsequent tables, a dash as an entry means that no
safe estimate for this quantity could be found. This occurs
particularly, but not exclusively, in cases like gcs004 or gtr121,
where the thin part of the bar is so weak that it is very hard to
measure its extent as reliably as we did for other simulations. We
will therefore omit these cases from most of the following
discussions, where the barlength estimate is necessary. Some values
for gtr101 and gcs006 are also unsafe because the barlens 
covers most of the thin bar extent and the barlens shapes are less
realistic (Sect.~\ref{sec:discussion}). We have, nevertheless, kept
some of these
value in the tables as a reminder that there are cases where the
barlens extent covers a very large fraction of that of the bar.   

The two most elongated
barlenses are those of gtr101 (axial ratio 0.43) and gtr102 (0.45), 
both cases being simulations with no gas and amongst the strongest bars
in AMR13. The remaining simulation with no gas (gcs006) has an axial
ratio of 0.64. Note that these simulations are the ones for which the
ratio of barlens to bar extent is the most unsafe. In all simulations
with gas, the  
barlenses are considerably nearer to near-circular, with most
axial ratios between 0.7 and 1.0. 
The ratio of the extent of the barlens to that of the thin bar component has
a lower limit of $\sim$0.4 and an upper one of $\sim$0.9. We will
discuss the  
implications of this result in Sect.~\ref{subsec:orbits}. 
 
Table \ref{tab:tellipsefits} gives information on the barlens
shape and on the ratio of its extent to that of the bar, for three
simulations and six snapshot times, spaced by 1 Gyr. These three simulations
are good examples of a very strong, an intermediate and a weak bar.
This table shows clearly that the
fractional change of the barlens shape, as well as of its extent
relative to the bar vary considerably from one case to
another. Amongst the three, the least
variation is found for gtr119, where the changes are barely of the
order of 10 per cent, and the most for gtr101 where the changes are
nearly of the order of 50 per cent. In general we can see that the
runs that showed the most secular evolution (see AMR13) show also the largest
changes of the barlens axial ratio and of its extent relative to that of
the bar, as expected. 

\liaedit {Simulation gtr111 has  
two distinct time intervals in its evolution with different properties
for the barlens in each. For times up to and including 6 Gyr the
barlens is near-circular and relatively short, with axial ratios
larger than 
0.9 and a length relative to the bar length of less than 0.4. It then
shows a clear change between 6 and 8 Gyr, while from 8 Gyr onward it
is considerably thinner and longer, with average axial ratio and
average relative extent of 0.75 and 0.6, respectively. It is
interesting to note that in the range between 6 and 8 Gyr  
the increase of the bar strength with time shows a change of slope,
turning from a rather fast to a more moderate growth (fig. 7 of AMR13
or fig. 1 of \citealt{Iannuzzi.Athanassoula.15}). In this same time
range, the increase of the peanut strength also changes from
rather fast to more moderate \citep[fig. 2
  of][]{Iannuzzi.Athanassoula.15}. This constitutes a further
argument for a strong link between the barlens and the bar. Note 
that no buckling episode (i.e. no asymmetry) can be seen by eye during
this time range, or at any other time. A more in depth analysis of this
interesting behaviour is beyond the scope of this paper.}   

\section{Decompositions: Technique and results}
\label{sec:decompositions}

\begin{figure*}
  \centering
  \includegraphics[scale=0.69]{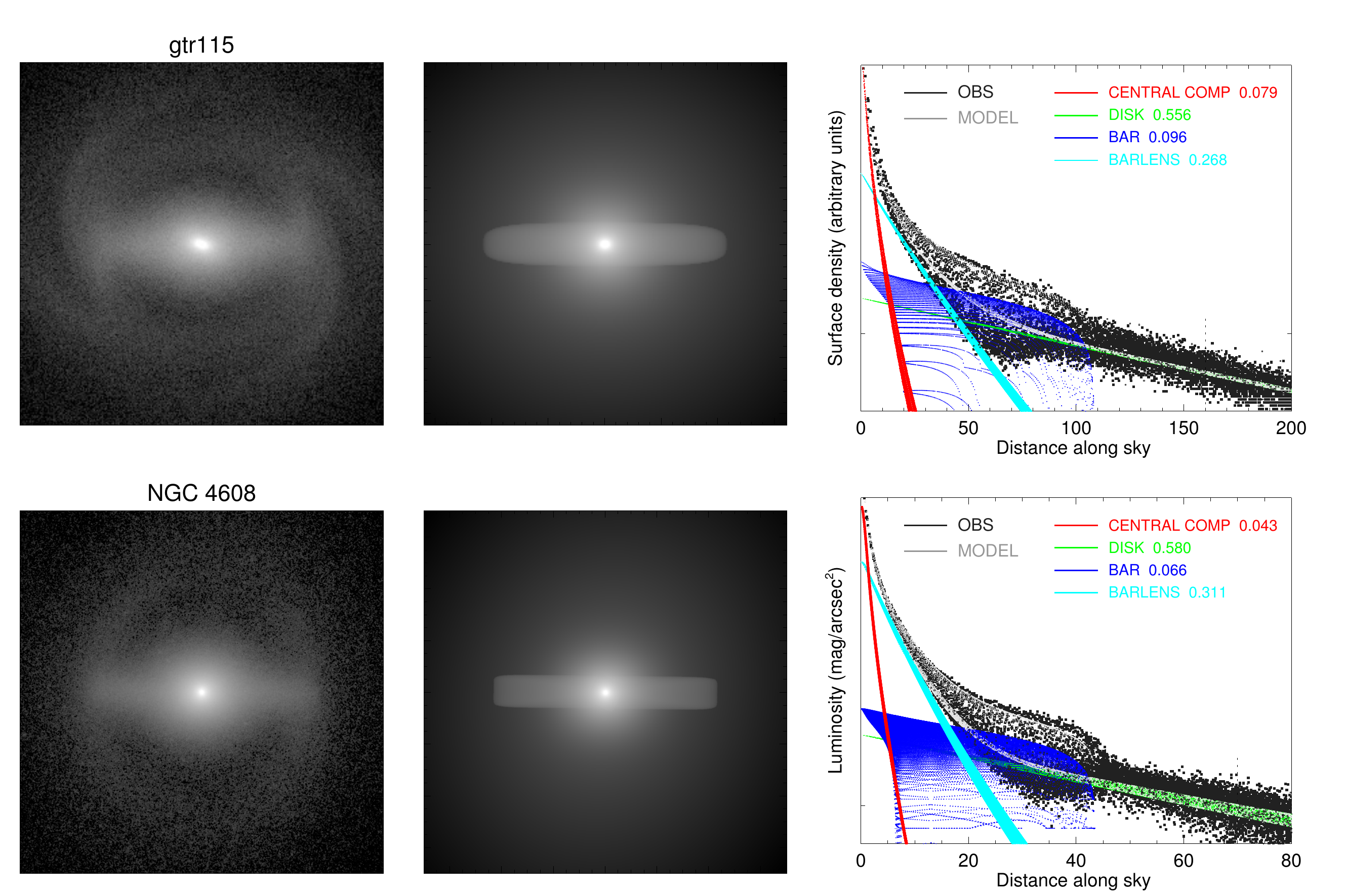}
  \includegraphics[scale=0.49]{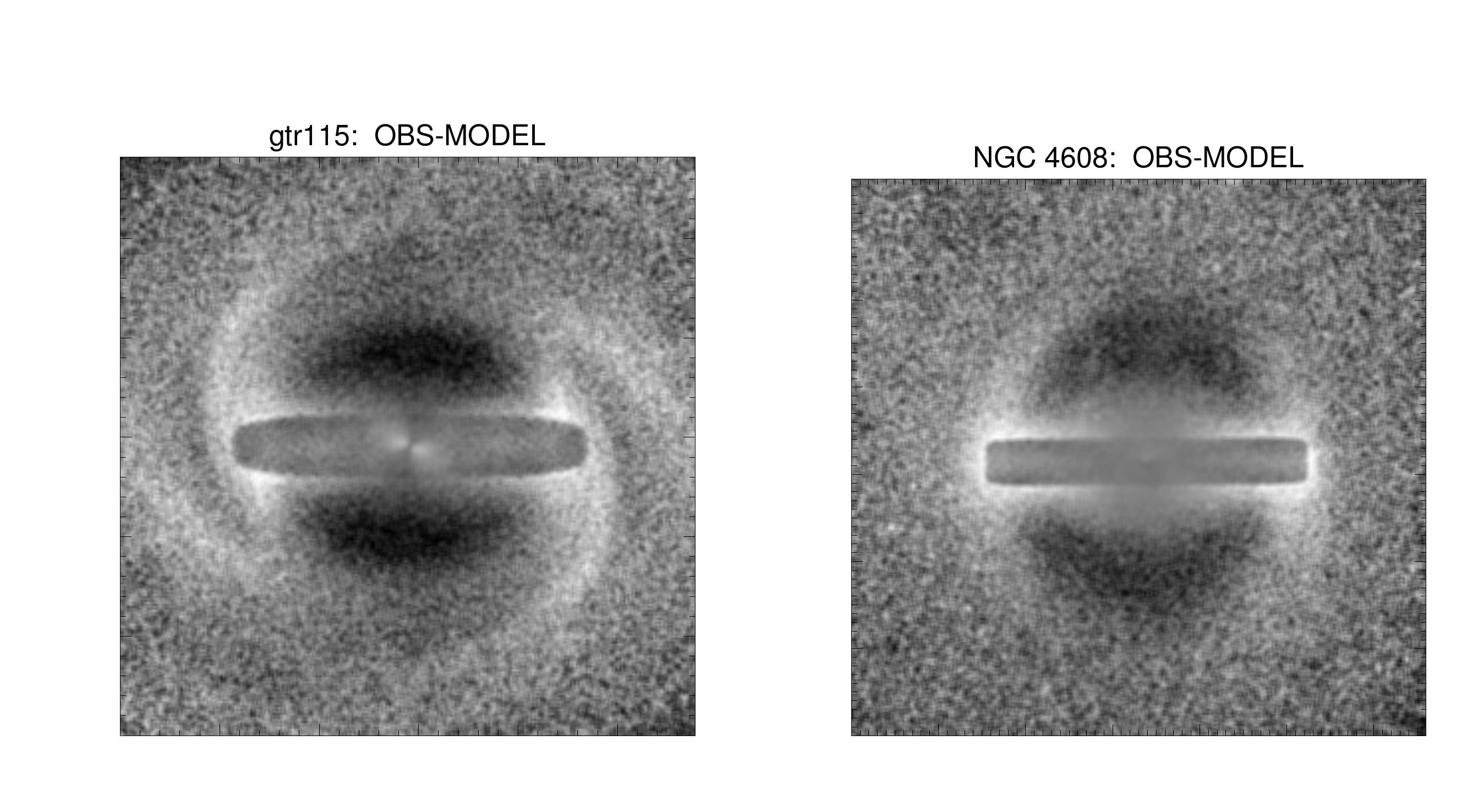}
  \caption{
    Images and their decompositions for the
  simulation model gtr115 (upper panels), and the galaxy NGC 4608
  observed in $K_s$-band (middle panels). The left-hand panels in the two
  upper rows show the
  images and the middle ones show the sum of the components obtained from
  their decomposition. All the images are shown in face-on view and
  cut so as to show only the bar and inner ring region. The right-hand
  panels in the two upper rows show two dimensional representations of the
  surface brightness profiles of the images (black dots), as well as
  of the GALFIT decomposition models fitted for the different
  structure components (red dots for the central peak component, green
  for the disc, 
  dark blue for the bar and cyan for the barlens). The total
  decomposition models are shown with light grey dots. `Bar' in the
  profiles refers to the thin, elongated bar
  component and `barlens' to the thick component, which is not far
  from circular. 
  The abscissa in the right, second-row panel is given in
  arcseconds. In the upper right panel the abscissa is
  scaled so that the bar length is roughly the same for the galaxy and
  for the simulation. The vertical black dotted lines in the profile
  figures (roughly at a distance from the centre of 160 and 70
  pixels for
  the upper and middle panels, respectively) show the edges of the
  images in the left and middle panels. 
  The two figures in the bottom panel show the 
  residuals for the simulation (left-hand panel) and for the galaxy
  (right-hand panel). In the dark areas the decomposition model has higher
  values than the simulation/galaxy, the opposite being true for the
  white ones. Note that no spirals or
  ring components were used in the decompositions, which accounts for
  the white or very light grey areas in the two lower panels. The
  method and the functions used for making the decompositions are
  described in the text.   
    }
\label{fig:comp-NGC5101_gtr115}
\end{figure*}

\subsection{The components}
\label{subsec:components}

The simulation models discussed in the previous section
were decomposed into multiple structural 
components, fitting the bar with two separate functions.  We used
GALFIT \citep{Peng.HIR.10} with the auxiliary illustration programs 
described in \cite{Salo.plus.15}, and the same approach as that used 
for the observed images in the NIRS0S Atlas \citep{Laurikainen.SBKC.10}. 
For the weighting of the simulation images a constant sigma-image was
used because this mimics well the typical uncertainty map for ground
based near-IR images, where the noise is dominated by the sky background.
All fitting functions have generalized elliptical isophotes
\citep{Athanassoula.MWPPLB.90}, so that:

\begin{equation}
(|x|/a)^{c} + (|y|/b)^{c}=1,
\end{equation}

\noindent where $c$ is a shape parameter and $a$ and $b$ are the major
and minor axes, respectively. We will denote their axial ratio by $q$
= $b/a$. An isophote is boxy when 
its shape parameter $c > $ 2, diamond-like when $c < $ 2, and elliptical when 
$c$ = 2 (see fig. 1 of \citealt{Athanassoula.MWPPLB.90}). For the radial projected surface density of the disc
component we used an exponential function:

\begin{equation}
\Sigma(r) = \Sigma_{0,d}~\exp[-(r/h_d)],
\end{equation}

\noindent where $\Sigma_{0,d}$ is the disc central surface density,
and $h_d$ is the disc radial scalelength. The thin 
bar component is described for $r \le r_{out}$ by a modified Ferrers
function: 

\begin{equation}
\Sigma(r) = \Sigma_{0,th}[1-(r/r_{out})^{2-\beta}]^\alpha,
\label{eq:modeif-ferrers}
\end{equation}

\noindent where $\Sigma_{0,th}$ is the central surface density of the
thin bar component, $r_{out}$ is its distance  from the centre to
its outer edge, $\beta$ gives its  
central slope, and $\alpha$ is a measure of the sharpness of its outer
truncation \citep{Ferrer.77}. For  $r > r_{out}$ the projected
  surface density is set equal to zero.

There is no preferred function for the barlens component. One
  possibility is the S\'ersic function:

\begin{equation}
\Sigma(r) = \Sigma_{0,bl} ~\exp[-(r/h_{bl})^{1/n}],
\end{equation}

\noindent where $\Sigma_{0,bl}$ and $h_{bl}$ are the barlens central
surface density and its scale parameter, respectively, and $n$ the
S\'ersic index. This does not have a cut-off at some outer radius. An
alternative is to use a modified Ferrers function
(equation~\ref{eq:modeif-ferrers}) which does have a cut-off,
but is otherwise less well suited. We tried both and adopted the one
which gave a better fit for each case.

We also include in our decompositions an extra S\'ersic profile to fit the
centre-most regions where there is typically a central peak (spike) in
the surface brightness profile, which we need to account for
in order to obtain a good fit to the barlens. Such a
component is also found in observed galaxies and is accounted for in
the same 
way, i.e. with a supplementary narrow S\'ersic profile. Simulations
necessarily have a softening which smooths out the central component.
However, the simulations we consider here are of high resolution, with
a softening of 50 pc. Similarly, the NIRS0S observations have a full
width half-maximum of the PSF of 1 arcsec, while for the S$^4$G
observations this is roughly 2 arcsec. For NGC 4314, shown in
Fig.~\ref{fig:NGC4314}, this gives 47 and 94 pc, for the two samples
respectively.
We thus see that the simulations and the NIRS0S
have very comparable high resolutions, while the S$^4$G sample has a
somewhat lower resolution. The above estimates   
argue that, for both simulations and
observations, this spike is real but, at the risk of erring on the
side of caution, we will not discuss its parameters further here
and, unless otherwise
mentioned, none of the flux values (be they that of the barlens, thin
bar, or that of the total flux used for normalization) 
include the contribution of the spike.

\begin{table}
   \caption[]{Flux ratios of the barlens component as obtained from
     the decompositions of snapshot images at $t$ = 6 Gyr. } 
\begin{tabular}{ccrlc}
\hline
 \noalign{\smallskip}
\noalign{\smallskip}
Run & Halo &  initial & F$_{bl}$/F$_{thin}$ &  F$_{bl}$/F$_{tot}$  \\
  &  &  gas [\%] &  &   \\
 \noalign{\smallskip}
 \noalign{\smallskip}
\hline
 \noalign{\smallskip}
\noalign{\smallskip}
gtr106  & 1: core/spherical &  20 &~~~~~2.6  & 0.33    \\    
gtr111  & 1: core/spherical &  50 &~~~~~0.93 & 0.13    \\   
gtr116  & 1: core/spherical &  75 & ~~~~~1.2  & 0.15    \\   
gtr119  & 1: core/spherical &  100 & ~~~~~5.8  & 0.13    \\  
 \noalign{\smallskip}
gtr109  & 2: core/mildly triaxial & 20 & ~~~~~3.1 & 0.39   \\
gtr114  & 2: core/mildly triaxial & 50 & ~~~~~3.5  & 0.25   \\ 
gtr117  & 2: core/mildly triaxial & 75 & ~~~~~0.4  & 0.26   \\
gtr120  & 2: core/mildly triaxial & 100 & ~~~~~9.5  & 0.19   \\
 \noalign{\smallskip}
gtr110   & 3: core/strongly triaxial & 20 & ~~~~~6.7 & 0.44   \\   
gtr115   & 3: core/strongly triaxial & 50 & ~~~~~1.9 & 0.19   \\    
gtr118   & 3: core/strongly triaxial & 75 & ~~~~~3.9 & 0.14   \\    
gtr121   & 3: core/strongly triaxial & 100 &~~~~~~~-   & 0.18   \\    
\noalign{\smallskip}
\noalign{\smallskip}
gcs001  & 4: cusp1/spherical & 20 &  ~~~~~~~-  & 0.21\\ 
gcs002  & 4: cusp1/spherical & 40 & ~~~~~2.9 & 0.18\\ 
gcs003  & 4: cusp1/spherical & 60 & ~~~~~3.5 & 0.20\\ 
\noalign{\smallskip}
gcs009  & 5: cusp2/spherical & 60 & ~~~~~2.3 & 0.18\\ 
gcs010  & 5: cusp2/spherical & 80 & ~~~~~4.4 & 0.21\\ 
 \noalign{\smallskip}
\hline
\end{tabular}
\label {tab:rdecompositions}
\end{table}

There are two important points to note in the above choices of the
components and of their properties. The first one is that we adopted
generalized ellipses, which is particularly crucial for the thin bar
and barlens components, because their isophotes do not have an
elliptical shape. The second important point is that we
describe the bar as a sum of two components. This is necessary
because, as was initially discussed by \cite{Athanassoula.05},
the shape of the bar is very complex, composed of two parts with very
different properties: an outer thin, and an inner thick component,
with different radial profiles. Such a component
does not exist in the GALFIT program, so we have to resort to using two
independent components to fit the bar adequately.  
Both components are crucial for the type of work we
need to do. Note that a second component for the bar has
been introduced in
some barred galaxy decompositions already about 10 years ago
\citep{Laurikainen.SB.05}. 
To our knowledge, however, it has not as yet been used for any
decompositions of simulation results.

\subsection{The fitting procedure}
\label{subsec:fit-procedure}

In our GALFIT decompositions we used the same procedure as in 
observational studies \citep{Laurikainen.SBKC.10}. The bar length was set to be 
the visually estimated outer edge of the bar obtained during the
ellipse fitting work (Sect.~\ref{sec:ellipsefits}).  This was
reasonable, because otherwise GALFIT would try to include the inner
ring as part of the bar. Furthermore, we fixed the barlens size and
  axial ratio to the values found from the ellipse fits. After that, we
started the fitting process, iterating in order to find the appropriate
values for the remaining parameters, and making a large number of
trials until a `final' decomposition 
was adopted. The choice was based on inspecting the profile fitting, 
and on a comparison of the decomposition model with the original 
image. Generally, it was not possible to get good fits by setting free
all parameters at once. We thus fixed some values and let GALFIT
search for the remaining. Once a reasonable value was found for one
parameter, it was fixed and others set free. These cycles were
continued until the final decomposition was reached. The number of
necessary trials varied from one image to another, but was always
large. Also the sequence in which we set the parameters free
varied from one 
case to another, although we often found easiest to obtain first
reasonable values for the orientations, the fluxes and the disc scalelength.  
 
\subsection{Results}
\label{subsec:decomp-results}

\subsubsection{Comparing decomposition results of an observed and a
  simulated galaxy}
\label{subsubsec:decomp-comparisons}

Fig.~\ref{fig:comp-NGC5101_gtr115} compares the decomposition results
for NGC 4608 from the NIRS0S sample and for simulation gtr115. The left-hand 
panels show the respective images and the middle ones the superposition 
of the components resulting from the decompositions. This comparison is made
more quantitative in the right-panels panels, which show the radial surface
density profiles of the various components separately, as well as the
totals. Such plots are 
standard in observational studies and are made so as to include
information from non-axisymmetric 2D decompositions in radial
profiles. Every black point is a pixel from the image and every red
(green, dark blue and cyan) point is a pixel from the central peak
component (disc, bar
and barlens component, respectively). Thus this plot can be seen as a
sequence of radial surface density profiles for each component
separately, the total model and the image. From all these plots we can
see that the simulation snapshot describes NGC 4608 quite
successfully, even 
though it was not made specifically for that purpose but is just a
closely resembling snapshot from the AMR13 sample. 

\liaedit {The lower panels of Fig.~\ref{fig:comp-NGC5101_gtr115} present
the residual maps, which can attest to the quality of the
decomposition in the relevant area. In our decomposition the disc
model is axisymmetric, so the minima on the direction of the bar minor
axis above and below the bar, seen both in the observed and the
simulated galaxy, can not
be accounted for. This leads to the dark areas in these regions.
Furthermore, no spiral or ring component
was included in the decomposition, since this is not the object of our
study. This explains the residuals outside the bar region where such
components are present in the simulation and galaxy. Note also that
the spirals in the outer parts of the 
simulation are not present in the galaxy image.  
This can be understood because the simulation 
has at the time of the comparison still nearly 6\% of its baryons in
the form of gas, which NGC 4608, being a SB0 without HI detection, 
does not have. Thus we can conclude that the decomposition is
satisfactory in the bar region under study, but further
components need to be introduced in order to get a better model for the whole
galaxy.}


\begin{table}
   \caption[]{Time evolution of the relative flux of the
     barlens component, as obtained from decompositions of the
     simulation images. } 
\begin{tabular}{lccccccc}
\hline
\noalign{\smallskip}
\noalign{\smallskip}
 Age  -$>$      & 4 & 5 & 6 & 7 & 8 & 9 & 10 \\
\noalign{\smallskip}
  (Gyr)         &   &   &   &   &   &   &    \\
\noalign{\smallskip}
\noalign{\smallskip}
\hline
\noalign{\smallskip}
\noalign{\smallskip}
\noalign{\smallskip}
{\it  gtr111:}& & & & & & & \\
\noalign{\smallskip}
 F$_{bl}$/F$_{thin}$ &0.44 &0.81 &0.95 &- &2.35 &1.39 & 1.36\\
 F$_{bl}$/F$_{tot}$  &0.03 &0.08 &0.13 &- &0.22 &0.20 &0.21 \\
\noalign{\smallskip}
{\it gtr119:} & & & & & & & \\
\noalign{\smallskip}
 F$_{bl}$/F$_{thin}$ &6.16 &6.28 &5.81 &2.87 &2.53 &1.79 &1.79 \\
 F$_{bl}$/F$_{tot}$  &0.11 &0.16 &0.16 &0.16 &0.17 & 0.20& 0.20\\
 \noalign{\smallskip}
\hline
 \noalign{\smallskip}

\end{tabular}
\label {tab:tdecompositions}
\end{table}

\subsubsection{Decomposition results for a number of simulations and times}
\label{subsubsec:decomp-results}

In Tables \ref{tab:rdecompositions} and \ref{tab:tdecompositions}
we give the decomposition results for the flux ratios of all our
models. The former has the
same layout as Table \ref{tab:rellipsefits}, except that in the fourth
column we give the values of the barlens fluxes, F$_{bl}$, normalized
to the flux of the thin bar component, F$_{thin}$, and in the 
fifth one the same quantity, now normalized to the total flux of the
galaxy, F$_{tot}$. Table~\ref{tab:tdecompositions} has a layout
similar to that of  
Table~\ref{tab:tellipsefits}, but gives information on the barlens flux
normalized by the flux of the thin bar component (upper line
for each simulation), and by the total flux (lower line). 

For simulations with no gas initially, e.g. gtr101, 
gtr102, gtr003 and gcs006, it is unclear whether 
a two bar component fit is necessary and decompositions with
a single bar component can not be excluded. We therefore do not
include these simulations in Tables \ref{tab:rdecompositions} and
\ref{tab:tdecompositions}.  
There are also decomposition difficulties
for snapshots at the other extreme of the gas
fraction, e.g. for gtr121, where the thin bar component is too weak
and short to allow a reliable decomposition. Simulation gtr117 was also difficult
to decompose and the values given are not as reliable as those of
other simulations. The cases
which are difficult to decompose presented also difficulties in the
ellipse fits. We will therefore single them out, or omit them from
many of the following 
discussions, since the results, if at all available, are less reliable
than those of the remaining runs.

In cases where
the thin bar component is very short and difficult to discern,
the F$_{bl}$/F$_{thin}$ takes very large values, but these are very
unsafe because of the large uncertainties for F$_{thin}$. They
should thus be used only with caution. This is presumably  
the case e.g. for the three earlier times of gtr119 in
Table~\ref{tab:tdecompositions}. 
For such cases, but also more generally, the
F$_{bl}$/F$_{tot}$ values are much safer to use. 

\subsubsection{Decomposition tests}

\label{subsubsec:test-decomp} 

\begin{table}
   \caption[]{The effect of the viewing angle on the
properties of barlenses for the simulation model gtr111 at time $t$=6
Gyr 
}
\begin{tabular}{ccccc}
\hline
 \noalign{\smallskip}
\noalign{\smallskip}
 Bar position  & inclination & b/a &  F$_{bl}$/F$_{thin}$ &  F$_{bl}$/F$_{tot}$  \\
 angle  &  &  & &   \\
 \noalign{\smallskip}
 \noalign{\smallskip}
\hline
 \noalign{\smallskip}
\noalign{\smallskip}
00 &00 &0.96 &0.95 &0.130 \\
00 &30 &0.77 &0.95 &0.110 \\
30 &30 &0.80 &0.99 &0.116 \\
60 &30 &0.75 &1.30 &0.135 \\
00 &60 & 0.64&0.50 &0.054 \\
30 &60 &0.78 &0.79 &0.089 \\
60 &60 &0.66 &0.31 & 0.058\\
90 &60 &0.86 &1.07  &0.118 \\
 \noalign{\smallskip}
 \noalign{\smallskip}
\hline
\end{tabular}
\label{tab:view-decomp}
\end{table}

We have so far considered only face-on views
with the bar major axis along the $x$ axis. Here we will present
decomposition results for different viewing angles.
For this we used simulation gtr111 at $t$=6 Gyr, whose barlens is
near-circular, with an axial ratio of 0.96.
We viewed the snapshot after a rotation in the disc plane, to put the
bar major axis at a different angles with respect to the galaxy major
axis. We then inclined it by another angle. The results are
given in Table~\ref{tab:view-decomp}. The first two columns give the
viewing angles, the third one gives the axial ratio of the barlens
component as obtained from the ellipse fits and the last two give the
relative barlens fluxes. 
The results for the face-on view are given in the first row, and
examples with 30$^{\circ}$ inclination in the next three rows. The errors
are of the order of 10--20 per cent, i.e. sufficiently small to
allow considerable analysis and statistics. 
If, however, we consider inclinations as high as 60$^{\circ}$ (four
last lines of Table~\ref{tab:view-decomp}), the
errors are higher. This argues that for some statistical
work involving barlenses it may be necessary to introduce a lower
inclination 
cut-off than the commonly used 60$^{\circ}$ or 65$^{\circ}$. 
How much the results are affected by a given inclination should
vary from one case to another and should in particular depend on the
extent of the barlens above and below the equatorial plane.

\section{Discussion}
\label{sec:discussion}

\subsection{Range of values for the barlens relative lengths and axial ratios}
\label{subsec:lengthrat-axrat}

\begin{figure}
  \includegraphics[scale=0.25]{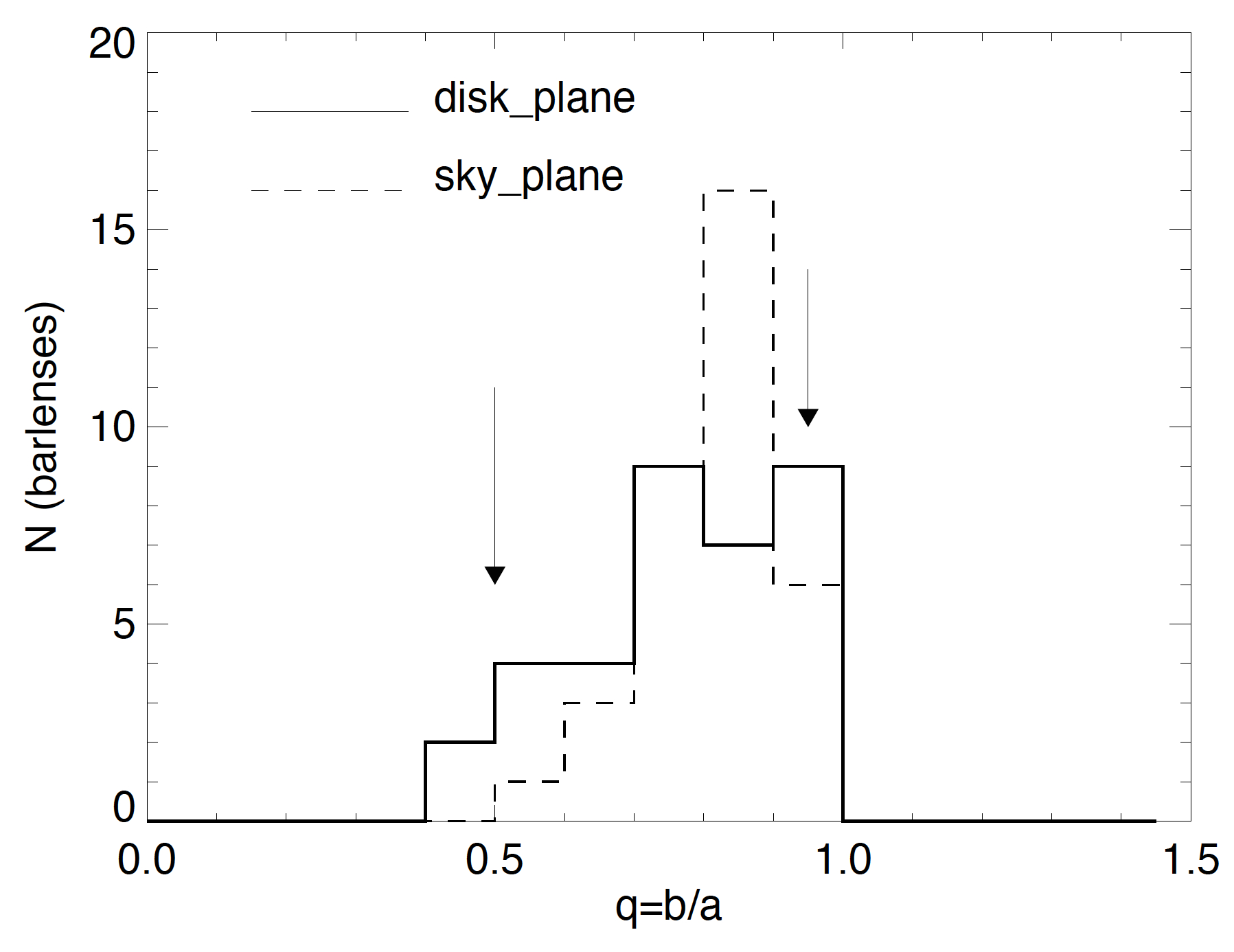}
  \caption{Number histogram of the minor-to-major
axis ratios of barlenses in the NIRS0S Atlas. The full black line
shows the measurements in the disc 
 plain, and the dashed line the same galaxies in
 the sky-plane.The arrows mark
the minimum and maximum of this parameter for our simulation models (in
the plane of the disc).}
\label{fig:qbarlens}
\end{figure}

\begin{figure}
  \includegraphics[scale=0.25]{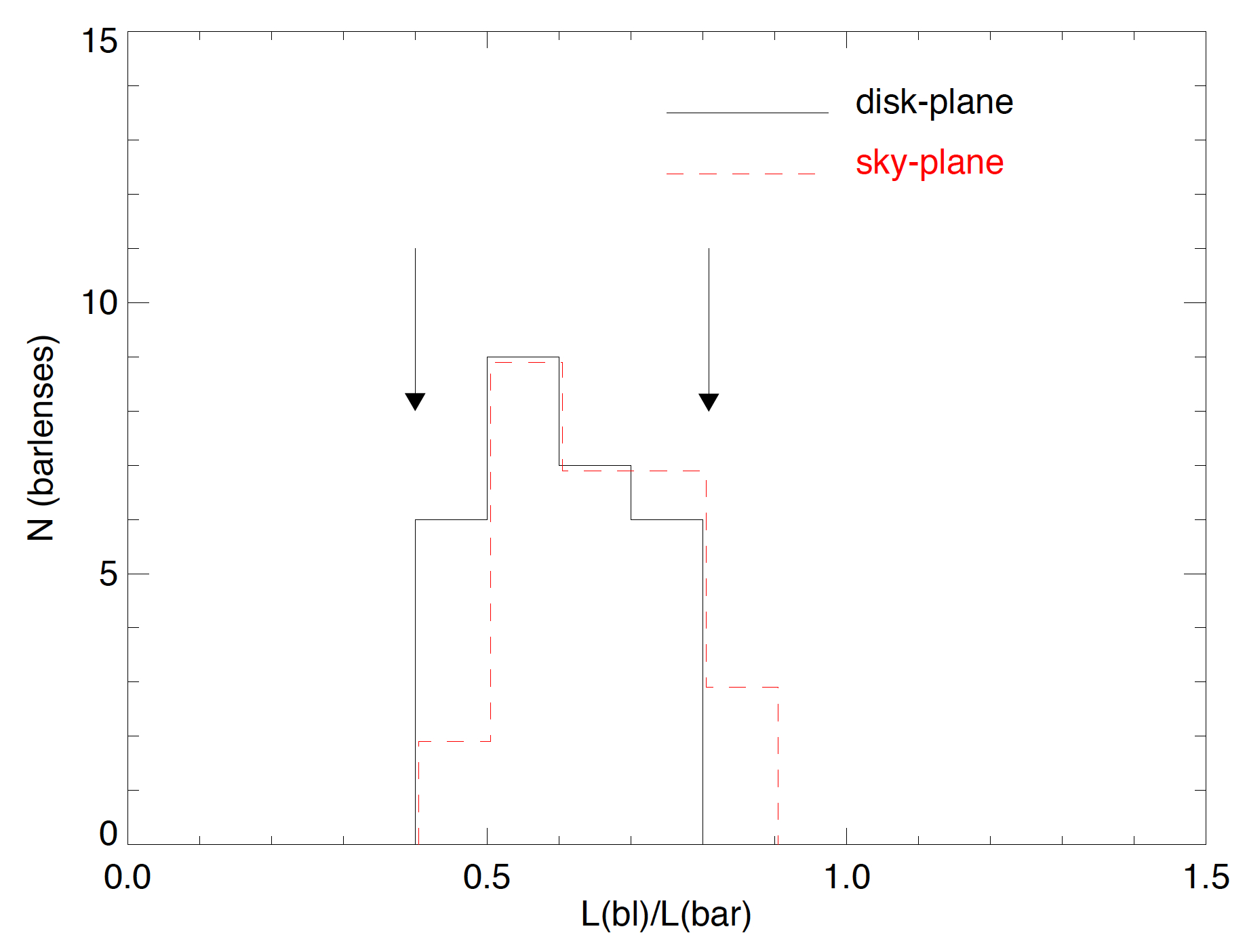}
\vskip 30pt 
  \caption{Number histogram of the sizes of barlenses for the galaxies
    in the NIRS0S Atlas, normalized to the corresponding
 bar length. The black full line shows the measurements in the disc
 plain, and the (red online) dashed line the same galaxies in
 the sky-plane. 
 The arrows indicate the minimum and maximum of the same parameter in
 our simulation models (in the disc plane). The sizes of structures
 are measured in a similar manner for the simulation models and for
 the observed galaxies.
    }
\label{fig:histoLb}
\end{figure}

Let us now compare the range of values for the axial ratios of the
barlens components in observations and in simulations. The former can
be found for all galaxies with a barlens component in Table 5 of 
\cite{Laurikainen.SBK.11} and their histogram is plotted here in  
Fig.~\ref{fig:qbarlens}, both in the sky-plane and in the galactic
disc plane (assuming thin
structures). The ellipticities for the simulations have been obtained
from the ellipse fits described in Sect.~\ref{sec:ellipsefits}. These
simulations, however, were made in order to understand the effect of
gas and of the dark matter halo on bar formation and evolution
(Sect.~\ref{sec:simulations}) and
do not necessarily cover the parameter space in the same way as the
barred galaxies in NIRS0S. Furthermore the deprojection of
observed galaxies, particularly barred, always entails some
uncertainty. It is thus more meaningful to compare the range of
allowed values than the distribution of values given by the whole histogram.   
We find that the range of ellipticities obtained for barlenses in our
simulation models (marked with arrows) fits well the range 
defined by the observations. Furthermore, barlens ellipticities of the
gtr models with initial gas fractions of 0.5, 0.75 or 1.00,
are within the range of 0.7--1.0, where most of the observed
ellipticities are found.

The distribution of barlens sizes, normalized to the
bar length, as obtained from the NIRS0S sample, are shown in
Fig.~\ref{fig:histoLb}, again given separately for the disc and the
sky planes. In the comparison with simulations we omitted gcs006,
because its measurements are much less reliable than those of the
other simulations (see Sect.~\ref{sec:ellipsefits} and
\ref{sec:decompositions}). Fig.~\ref{fig:histoLb} shows clearly the
comparison of the range of 
values obtained from simulations to the range of observed ones  
is most satisfactory. 

\subsection{Comparing the face-on and the side-on views}
\label{subsec:sideon}

\begin{figure*}
  \includegraphics[scale=0.7, angle=-90]{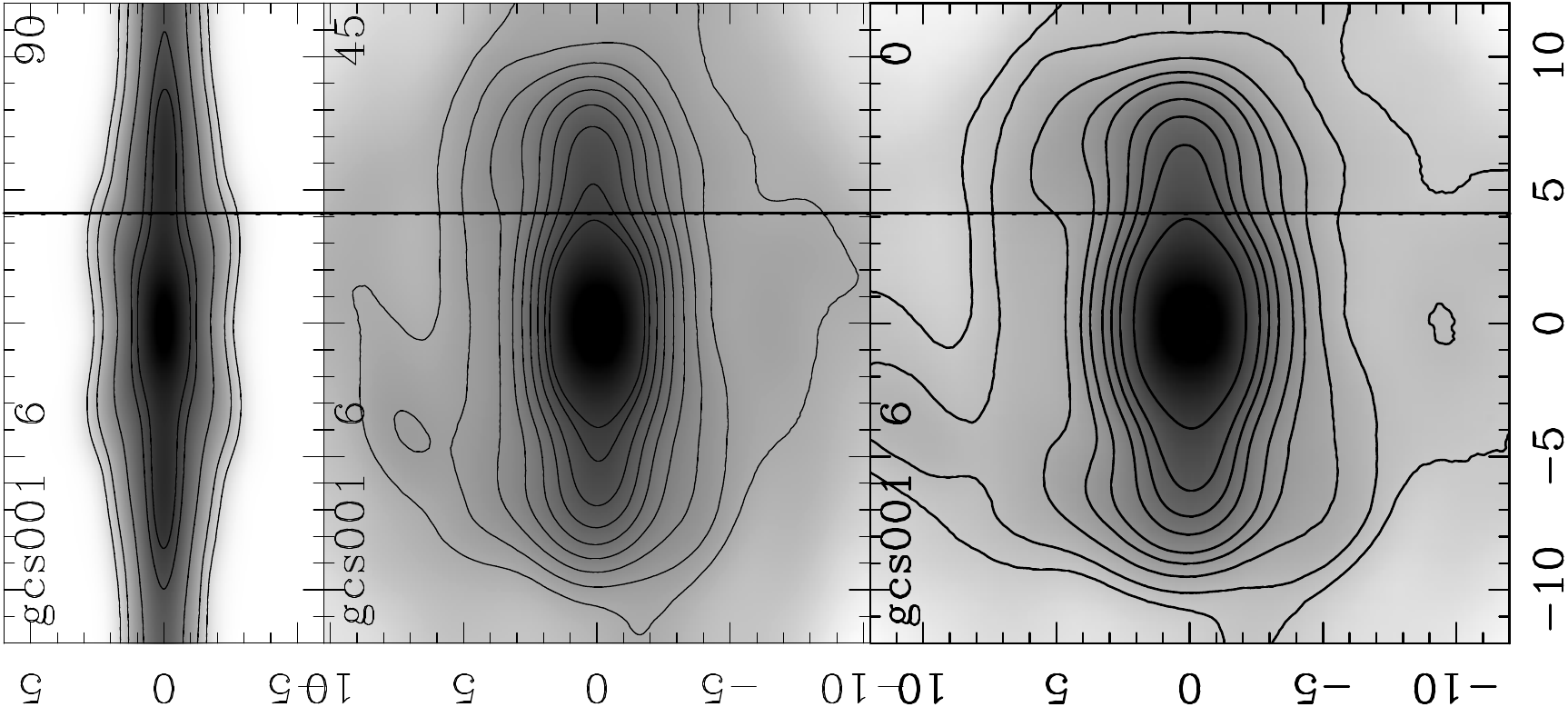}
  \includegraphics[scale=0.7, angle=-90]{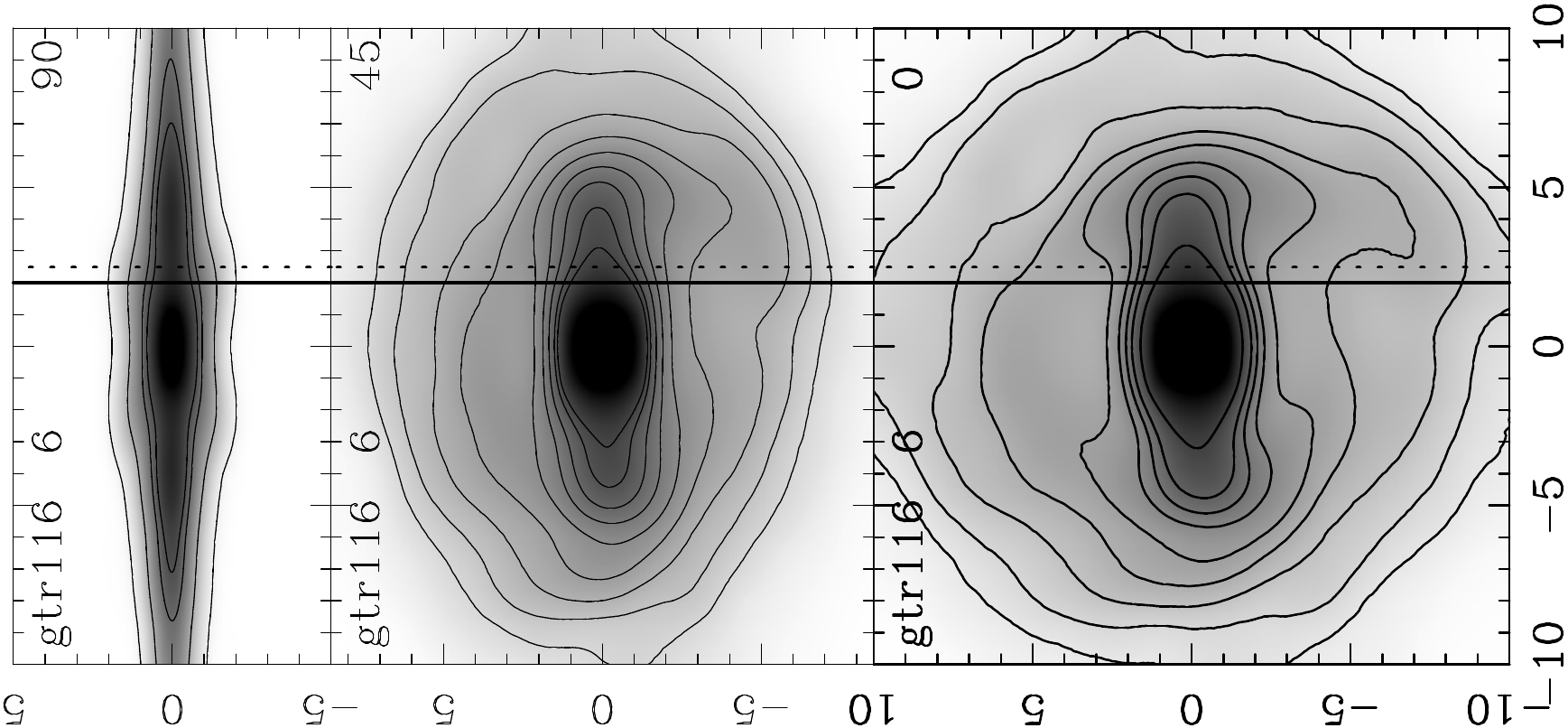}
  \includegraphics[scale=0.7, angle=-90]{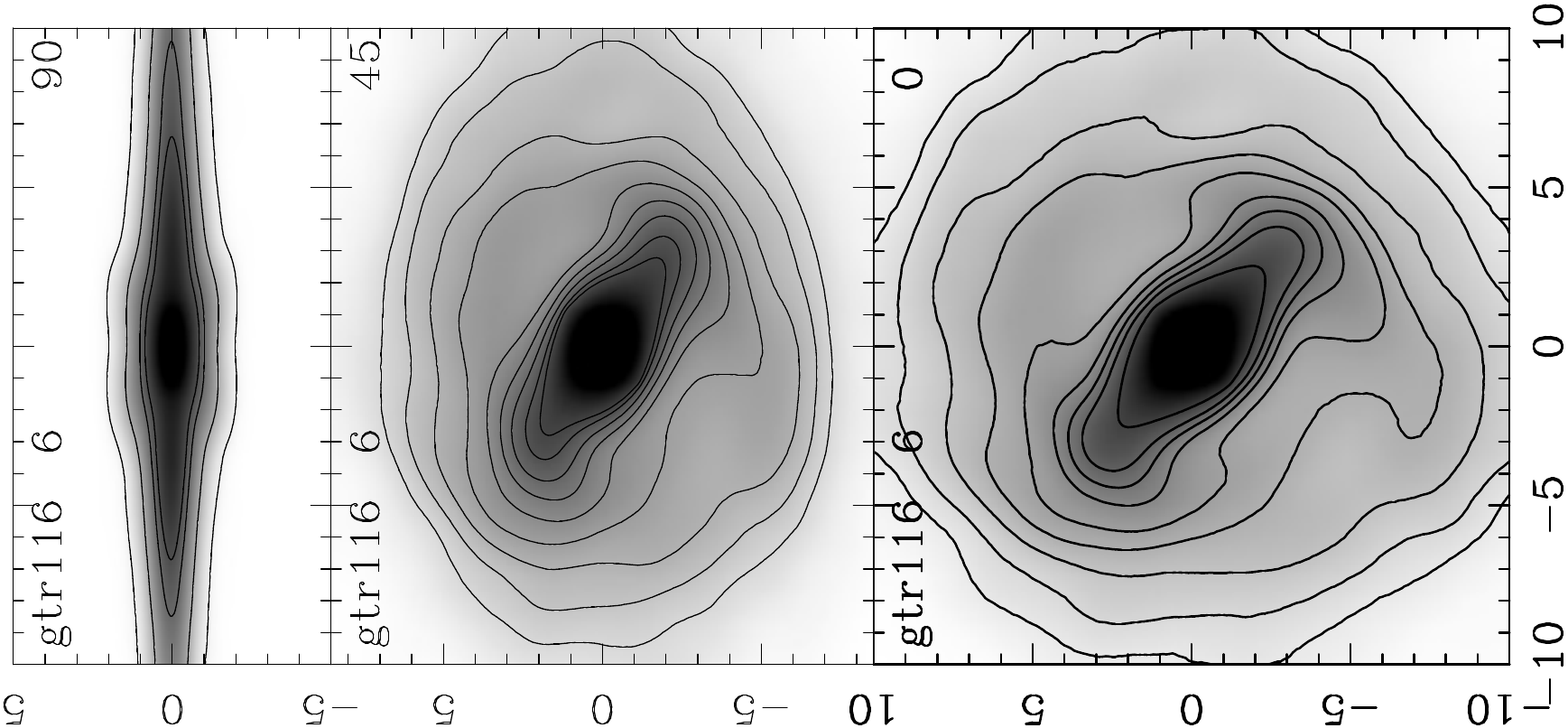}
  \caption{The left-hand and middle columns show grey-scale plots and
    isodensities of three views of simulation gcs001  
    (left) and gtr116 (middle), both at $t$=6 Gyr. In the
    lowest sub-panels the simulations are viewed face-on, in the top ones 
    side-on and in the central ones at $45^{\circ}$. The grey-scale levels
    are logarithmically spaced and the isodensities are at levels
    chosen to show best given morphological features discussed in the
    text. The vertical solid lines extending through all three 
    sub-panels show the extent of the barlens component in all
    three views, 
    as measured by the ellipse fits (Sect.~\ref{sec:ellipsefits}). 
    The vertical dashed lines show the extent of the B/P/X bulge. The
    right-hand column illustrates again gtr116, but now with the bar at
    45$^\circ$ from the line of nodes. In
    the upper left corner of each sub-panel we give the simulation
    name and snapshot time in Gyr and in
    the upper right one the viewing inclination angle in degrees.     
}
\label{fig:many-views}
\end{figure*}

Simulations have an important advantage over observations in that they
allow the viewing of an object from any desired viewing angle. We will 
use this here in order to study the 3D structure of the barlens 
component. In the left-hand and middle panels of Fig.~\ref{fig:many-views}
we show two snapshots viewed 
from three different angles. The snapshots correspond to the disc
component of simulations gtr116 and gcs001, both at $t$=6 Gyr.
In the lower sub-panels they are viewed face-on
and in the upper ones edge-on. The central sub-panels correspond to an 
intermediate viewing angle of $45^{\circ}$. In both cases the bar is 
oriented along the $x$ axis, to facilitate comparisons. These two
snapshots were chosen because 
they are good examples of specific morphological features which we
wish to discuss here. 

The face-on and intermediate views in Fig.~\ref{fig:many-views} show
that the edge of the barlens component at and around the direction of
the bar minor axis is relatively sharply defined, as witnessed by the
crowding 
of the isodensities in that region, arguing for a correspondingly
  sharp decrease of the barlens density profile. This is in
good agreement with the strong decrease of the surface density in the
barlens region seen on the minor axis profiles in
Fig. \ref{fig:comp-morpho}. 
In contrast, the only crowding of isodensities in the direction of the
bar major axis is at the end of the bar. No corresponding feature is
seen at the end of the barlens, in good agreement with
Fig.~\ref{fig:NGC5101_gtr115_prof} 
and the corresponding discussion at the end of
Sect.~\ref{sec:global}.
   
A further interesting point to note is the difference that the
inclination makes to the shapes of both the thin bar component and
the barlens. This can be clearly seen by comparing the shape of the
isodensities in the face-on and the intermediate view. In the face-on
view, the thickness of the bar along the direction of its minor axis 
has a maximum at $x$=0, then decreases with increasing
distance from the centre to reach a minimum before increasing
again in the region of the ansae. For example for gcs001 at $t$=6 Gyr
(left-hand panel of Fig.~\ref{fig:many-views}), the  
minimum thickness of the projected surface density isocontours occurs
somewhere 
around 4 kpc from the centre (measured along the bar major axis).  Then the bar
outline starts becoming thicker again. 
Such a minimum is not visible in the intermediate view (inclination angle of
$45^{\circ}$), where the barlens isophotes near the bar minor axis
are flat, presumably due to the projected contribution
of the B/P bulge which extends more than a kpc above the
equatorial plane. Similar results can be seen in the middle panel for
gtr116 at $t$=6 Gyr.

The solid vertical lines extending through all three sub-panels show
the extent of the barlens along the bar major axis, 
as measured from the ellipse fits (Sect.~\ref{sec:ellipsefits}).
There are also vertical dotted lines, which show the
horizontal extent of the B/P bulge as estimated from the
side-on view. Comparing their locations to that of the solid lines
it is clear that the end of the barlens coincides, to within the
estimation 
accuracy, with the end of the B/P bulge. In fact in some
cases, as e.g. gcs001 at $t$=6 Gyr (left-hand panel of
Fig.~\ref{fig:many-views}), the two estimates coincide to within the
thickness of the lines in the plot, so that no dotted line can be
seen. We repeated this 
exercise for a number of snapshots (simulations and times) and always 
found the same qualitative result. For the ratios of the extent of the
barlens to that of the B/P bulge we found a median of 0.96, a
mean of 0.97 and an absolute deviation of 0.14, which argues that the
barlens and the B/P  extent are equal to within the measuring errors.
This strongly argues that the 
barlens and the boxy/peanut bulge are one and the same component, and 
not two separate components, or two separate parts of the bar. They
are simply viewed from a different viewing angle, near face-on for the
barlens and near edge-on for the B/P/X bulge. One would then expect
that the fraction of observed galaxies having either a
barlens or a B/P/X should be independent of the viewing angle, and this
is indeed what we found and reported in the accompanying observational
paper (L+14). 

\begin{figure*}
  \includegraphics[scale=0.4, angle=0]{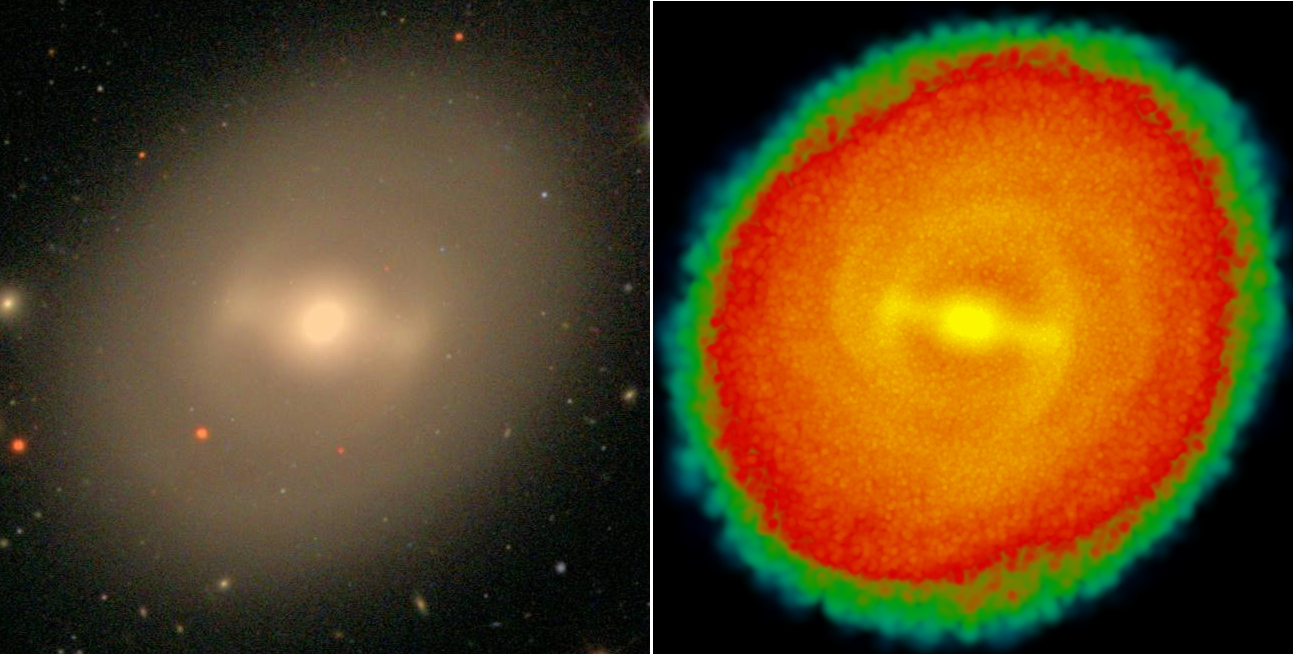}
  \caption{Comparison of a simulation snapshot (right-hand panel) with an
    SDSS image of NGC 936 (Hogg, Blanton and the SDSS collaboration,
    as in the Near Extragalactic Database). Note the resemblance
    between the two morphologies in the bar region. The simulation 
    has, by construction, no classical bulge component, the
    inner component being a barlens when viewed face-on and a
    B/P/X bulge when seen edge-on. We discuss the implication of
    this comparison in Sect.~\ref{subsec:NGC936}.     
}
\label{fig:NGC936}
\end{figure*}

We thus confirm that, as argued by \cite{Athanassoula.05}, bars are
composed of two parts, an outer one 
being both vertically and horizontally thin, so that it can be called 
the thin bar component, and an inner one which is vertically thick and 
also fatter in the
equatorial plane. This can be called the thick part of the bar, or the
B/P/X bulge, or the barlens; all three names describing the same part
of the bar. 

Our plots allowed us to obtain a rule of thumb that can give us
information on the extent of the B/P/X/barlens component when the
galaxy is viewed
near-face-on. Indeed, as already mentioned, the shape of the
isophotes in the barlens region is very different from that in the
thin bar part. This makes the isophotes change curvature, from
orienting their concave part towards the bar major axis to being rather
flat, or orienting their convex part towards the bar major axis. The
point or region where this change of curvature occurs gives a good
estimate of the length of the barlens component. When applying this
rule of thumb, it is also
recommended to use several isodensity contours, since some show this
minimum or flattening clearer than others.

This rule of thumb
works well near the face-on geometry, but it needs to be
extended to be applicable to inclination angles up to say 
45$^{\circ}$. Fig.~\ref{fig:many-views} shows that for such
inclinations and when the bar is near the line of nodes, the
isodensities near the bar minor axis are not curved 
as for the face-on view, but flat (straight). Thus 
the bar isodensities have two flat sections, both roughly parallel to
the bar major axis. The location where any one of these two sections joins
the part of the isodensity that orients its concave part towards the
bar major axis is the
end of the barlens. As before, it is more reliable to use a number of 
isophotes. It is not necessary to extend this rule of thumb to
yet higher inclinations (i.e. nearer to edge-on) because there the
vertical extent of the barlens makes it anyway easier to 
find its radial
extent \citep{Athanassoula.Beaton.06, Erwin.Debattista.13}. 

We checked our rule of thumb by applying it to a number of snapshots
seen from different viewing angles, and found that it fares very well
for most morphologies,
particularly after one's eye has gained some expertise in detecting the
correct features. There may, nevertheless, be some bar morphologies to 
which it can not be applied, such as relatively fat bars with
near-elliptical isodensities. Our rule of thumb 
is of course rather crude, but certainly safer than using a constant
value for the ratio of the bar and barlens extents, as has been previously
proposed, because as we saw in Sect.~\ref{subsec:lengthrat-axrat} and
will further discuss in Sect.~\ref{subsec:orbits} the
ratio of these two extents covers a wide range of values both in the
observations and  in the simulations.

\subsection{Can some barlenses be partly or fully mistaken for classical bulges?}
\label{subsec:NGC936}

In Fig.~\ref{fig:NGC936} we view a further comparison between simulations
and observations. The left-hand panel shows an image of the barred galaxy
NGC 936, obtained by combining the g, r and i images from the SDSS. The
right-hand panel displays  
gtr116 at time $t$ = 6 Gyr, viewed from angles such
that it mimics best NGC 936. This simulation was not
specifically run so as to model NGC 936, yet the bar regions in the
observations and in the simulations have a very similar
morphology. The most important thing to note, however, is that this
simple visual inspection of the central 
region of both NGC 936 and the simulation could lead to the conclusion
that both have a classical bulge, while at least for the simulation we
are sure, by construction, that it has no such component.
This shows that a barlens component may be mistaken
for a classical bulge, if morphology is the only available information.

Could photometry reveal whether the inner component is a barlens or a
classical bulge? Fig.  
\ref{fig:NGC5101_gtr115_prof} shows that, both  
in observations and in simulations, barlenses can produce a clear 
central peak in the radial luminosity profile. This, however, is less
sharp than that of a classical bulge, and quantitative measurements of
the S\'ersic index should be able to distinguish between the two, since the
standard value for a classical bulge is larger than 2 or 2.5
(\citealt{Kormendy.Kennicutt.04}, \citealt{Drory.Fisher.07,
  Fisher.Drory.08}), while that of the barlens is considerably smaller (L+14). 
Thus, the answer to our initial question is that accurate
photometry should most probably be able to distinguish between a
classical bulge and a barlens component. 

Could colours, or population synthesis help us distinguish between
these two types of components? Indeed, B/P/X bulges are 
formed from the vertical instabilities of the bar, i.e. such bulges
are constituted of stars initially in the disc. In contrast, the most
standard formation 
mechanism of classical bulges is from mergings occurring early on in
the galaxy formation process, before the formation of the thin disc.
Thus if the inner component is composed of stars that are older and
redder than the disc or bar, then 
it would be a classical bulge and not a barlens. This colour
difference, however, is not a
necessary condition, since classical bulges can also form from later
occurring minor mergers or they may have later gas accretion
\citep[e.g.][]{Aguerri.BP.01, Coelho.Gadotti.11}, in which cases their
stellar populations 
could be younger and bluer than the disc, or they could include a
younger sub-population. Thus, in some cases it may be possible to
distinguish classical bulges from barlenses by their colours or with
the help of population synthesis, but in others not.  

Orbits in barlenses are very different from those in classical bulges,
so detailed kinematic observations could be very helpful in
distinguishing between the two types of components. 
Specifically for NGC 936, signatures of kinematics which are not
compatible with pure classical bulges have been discussed by  
\cite{Kormendy.83, Kormendy.84, Kent.Glaudel.89} and Cappellari (2013, 
private communication, see  
www.eso.org/sci/meetings/2013/MORPH/Videos/Day-4/morph2013$_{}$Athanassoula.mp4;
see also \citealt{Cappellari.p.11} and \citealt{Cappellari.p.13}).
Unfortunately though, such detailed kinematics are available only for
relatively few galaxies\footnote{Three more near-face-on galaxies with B/P/X
  bulges have been observed by \cite{Mendez.ACDDRAP.08} and 
  more are included in the SAURON and ATLAS3D samples.}. 

All the above argue that, for galaxies which are seen near to face-on
and for which neither 
detailed kinematics nor photometry are available, barlenses could
be mistaken for classical bulges. The picture is yet more
complex because B/P/X/bl components and classical bulges 
can well coexist in galaxies \citep[e.g.][]{Athanassoula.05, 
Kormendy.Barentine.10, Nowak.TESBD.10, Mendez.ADCA.14, Erwin.plus.15}. In such
cases, as discussed in
the accompanying observational paper (L+14), 
decompositions with only three components -- a disc, a classical bulge
and a bar -- will overestimate the bulge mass to try and compensate
for the omitted barlens. This effect can be very important (on average
a factor of 3.5) as could be
expected since the barlens can constitute a considerable part of the
total mass, as shown here for simulations (see Tables
\ref{tab:rdecompositions} and \ref{tab:tdecompositions},   
and Sect.~\ref{subsec:trends}) and in (L+14) for observations.

Mistaking a barlens, or part thereof, for a classical bulge can have
repercussions for studies of individual galaxies. For example, in 
order to calculate potentials and forces within a barred galaxy in view of
studying its orbital structure or the gas flow in it, we need first to
obtain the volume density from the projected surface density. Here the
difference 
between the vertical distribution of mass in barlenses and in classical bulges
can make non-negligible differences in the results. Similarly, it can
affect the calculation of the $Q_b$ measure of the bar strength. 
Moreover, such mistakes may bias statistical studies involving
galaxy classifications, and erroneously move galaxies
towards earlier types. 
It is clear that the
result of such misclassifications is a bias and not a statistical  error,
so that the correct fraction of disc galaxies with no
classical bulge is in reality larger than so far estimated, and the 
contribution of classical bulges to the total baryonic mass budget considerably
less. This is in good agreement with recent detailed studies which
have argued that also in our Galaxy \citep{Shen.RKHDK.10, Ness.p.13a,
  Ness.p.13b}, as well as in a number of nearby 
Sc - Scd galaxies \citep{Kormendy.DBC.10} the classical
bulge component is either non-existent or very small. Here we extend
this statement to early type barred galaxies.
A smaller contribution of classical bulges to the total baryonic mass
fraction and a larger number of disc galaxies with no classical bulge
are facts that future galaxy formation theories will
have to take into account.  

\subsection{Trends}
\label{subsec:trends}

\begin{figure}
  \includegraphics[scale=0.65, angle=-90]{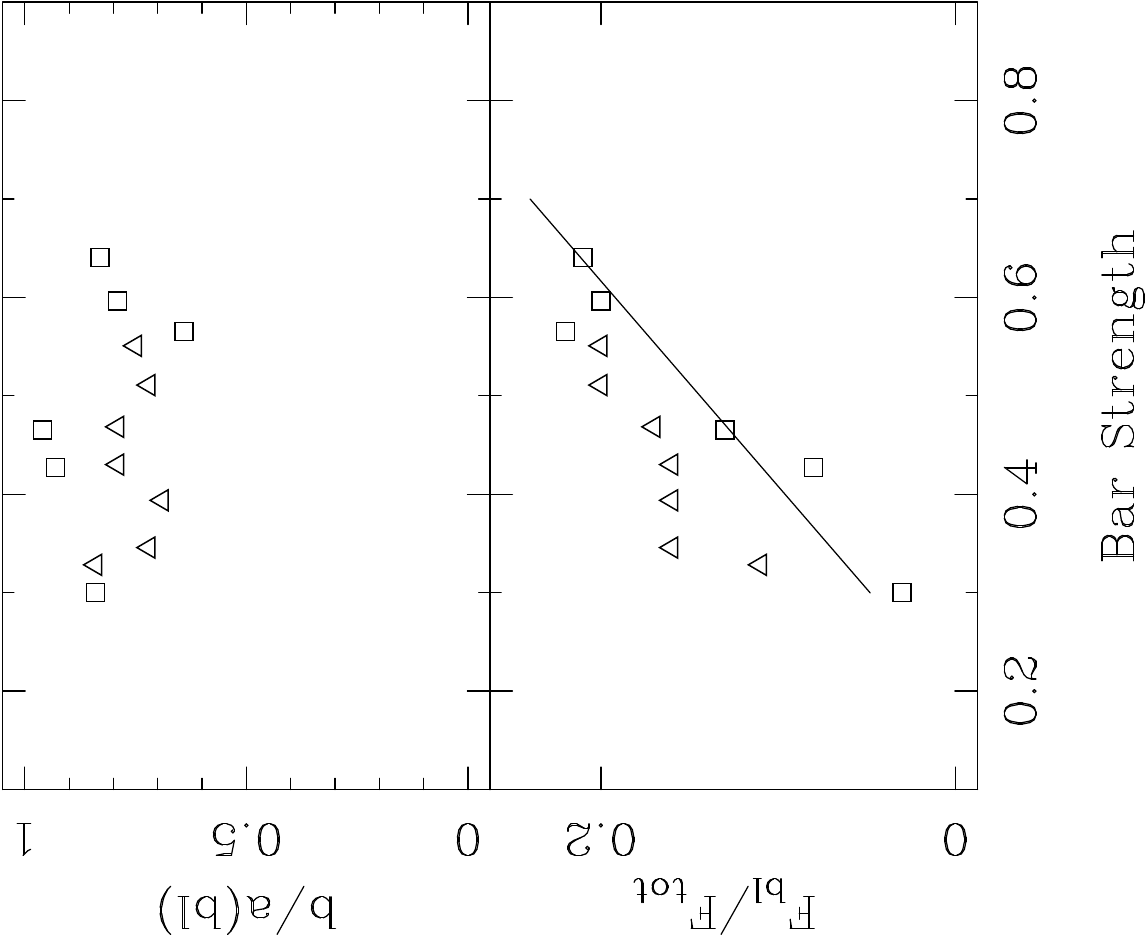}
  \caption{Evolutionary trends of barlens parameters. We display barlens axial ratio
    (upper panel) and barlens flux normalized by the total flux 
    (lower panel), both as a function of bar strength. Data from
    gtr111 are plotted with open squares and data from gtr119 by open
    triangles. As the bar strength increases with time, so does the
    relative importance of the barlens, amounting to a bigger
    fraction of the total baryonic mass. The solid line is the regression
    line obtained from the observations (L+14).  
    For more information see Sect.~\ref{subsec:trends}.
}
\label{fig:trends}
\end{figure}

In this section we will use the data we obtained in
Sects.~\ref{sec:ellipsefits} and 
\ref{sec:decompositions} to follow the temporal
evolution of the barlens properties. In
principle we could also check the effect of the gas fraction on the
barlens, but such information is more difficult to
interpret. Indeed at any given time the different simulations are at
different stage of evolution, because the bar formation time
depends on the gas fraction (AMR13).  

In Fig.~\ref{fig:trends} we show the evolution of barlens properties
as a function of time for simulations gtr111 and gtr119. In order to
allow comparisons with observations we display all quantities as a
function of bar strength, which is an observable
quantity. 
How the bar strength increases with time depends strongly both on the
fraction of gas in the disc component and on the halo triaxiality,
being the strongest for simulations with no gas and a spherical halo,
and weakest for simulations with a high gas fraction and a high halo
triaxiality (AMR13). Thus, although the
bar strength is not, strictly speaking, a substitute for time, it
is a monotonically (though not linearly) increasing function of 
time\footnote{This is true for the times 
in which our analysis is made, which are
  after peanut formation, i.e. in the secular evolution phase. At
  earlier times and particularly during a buckling, the bar strength is
  not a monotonic function of time \citep[see][for
    reviews]{Athanassoula.13, Athanassoula.15}. See also
  the discussion in Sect.~\ref{subsec:orbits}.}. The bar strength was 
calculated in AMR13, from the maximum of the relative $m$=2
Fourier component.
   
The upper panel of  Fig.~\ref{fig:trends} shows the barlens axial
ratio -- obtained from the 
ellipse fits described in Sect.~\ref{sec:ellipsefits} (Table
\ref{tab:tellipsefits}) -- as a function of the bar strength. No trend
is visible for either gtr111 or gtr119. On the contrary, gtr101, a
gas-less simulation with a spherical halo, has a very strong bar and
shows a trend, with the barlens becoming more elongated with time (not
shown here, but see values in Table~\ref{tab:tellipsefits}). However, as we
already stated, barlenses in gas-less 
simulations are less realistic than those in the remaining runs, and
only few galaxies have such elongated barlens shapes (see also
Sect.~\ref{subsec:lengthrat-axrat}). For gtr111 and gtr119 there is no
trend and the axial ratio values are in the range of 0.65 to 0.95,
in good agreement with the observations (Fig.~\ref{fig:qbarlens}). 
Thus our simulations argue that there should be no
clear evolution of the barlens shape, except perhaps for the strongest
bars.       

The lower panel shows the flux of the barlens component normalized  
by the total flux, as a function of bar strength. 
Simulations gtr111 and gtr119 show a clear evolution of their bar
strength (AMR13) as well as of their relative barlens flux, such that there is
a clear trend between these two quantities. Each
simulation separately gives a much tighter fit to a straight line than
the two together, 
because the points from the more gas-rich simulation (gtr119) are
displaced above those of gtr111. Furthermore, the slopes show that the
evolution of the barlens is stronger for the galaxy with less gas
(gtr111). We have thus shown that the fraction of the total baryonic mass
which is included in the barlens component increases noticeably with time, more
so in gas-poor cases.   

This global increase with time can be understood from the general growth
of the bar during the secular evolution phase \citep[][for a
  review]{Athanassoula.13}, which, as we show here, is followed by an increase of
the mass of its barlens component, in good agreement with the known
correlation between bar and peanut strength \citep{Athanassoula.08}.
Furthermore, the fact that the evolution for the gas poor simulation is
stronger than for the gas-rich one is also in agreement with the results of
AMR13, where bars in gas-poor discs were shown to evolve faster than bars
in gas-rich discs.   

Should we expect our predicted trends to be verified by observed galaxies?
This is well possible, but not necessary. It should not 
be forgotten that observations may contain a much more diverse 
set of galaxy parameters than our models, with different 
masses, different disc velocity dispersions, cases with and without
classical bulges, and/or a wider variety of halo mass distributions,
etc, which could `dilute' trends. Furthermore, simulations may
cover parts of parameter space which are not relevant to real
galaxies, e.g. have more extreme gas fractions or different halo shapes.

The above
question -- i.e. whether real galaxies verify our predicted trends -- was
answered in the accompanying observational paper (L+14)
using a selected sub-sample of 10--15
galaxies from the NIRS0S and S$^4$G samples. These show no
trend for the barlens axial ratios and a very clear trend for the
barlens relative flux. It is gratifying to note
that these observational results confirm our predictions. 

To extend the comparison further, we plotted the regression line from 
L+14 
in the lower panel of
Fig.~\ref{fig:trends}. This fits very well the data from gtr111, while
those of gtr119 trace a line of a somewhat smaller slope, and displaced
upwards. This shows that gtr111 represents better the observations
than gtr119. The main difference between the initial conditions of
these two simulations is the fraction of gas in the disc component,
which in the beginning of the simulation is 100\% for gtr119  and 50\%
for gtr111. Indeed it is more realistic to assume that at the time
when the disc got in place the gas fraction in it was of the
order of 50\% rather than 100\% \citep[e.g.][and references
  therein]{Genzel.p.14}. 
 
\subsection{Barlens formation}
\label{formation}

\begin{figure*}
  \includegraphics[scale=0.46, angle=-90]{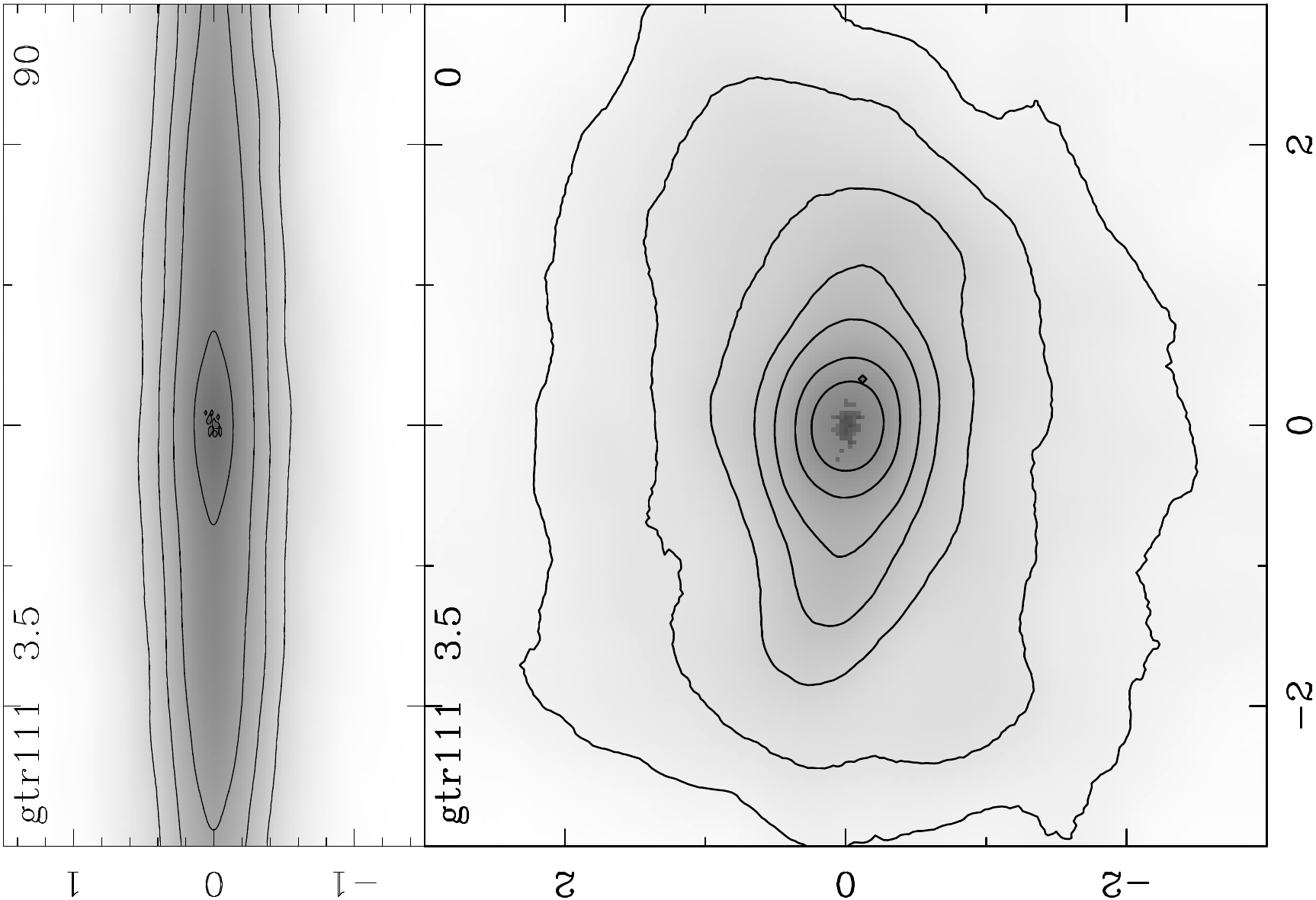}
  \includegraphics[scale=0.46, angle=-90]{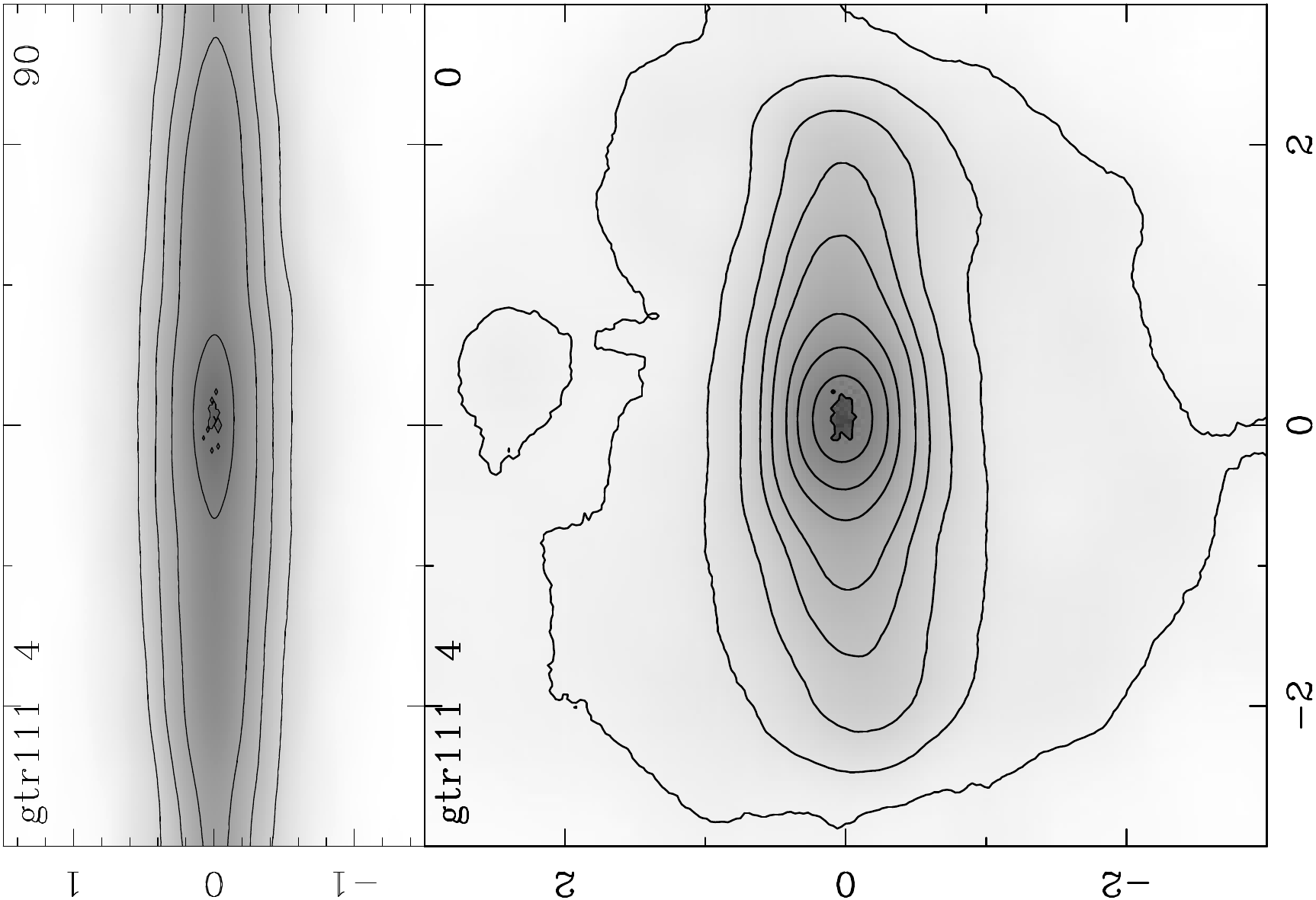}
  \includegraphics[scale=0.46, angle=-90]{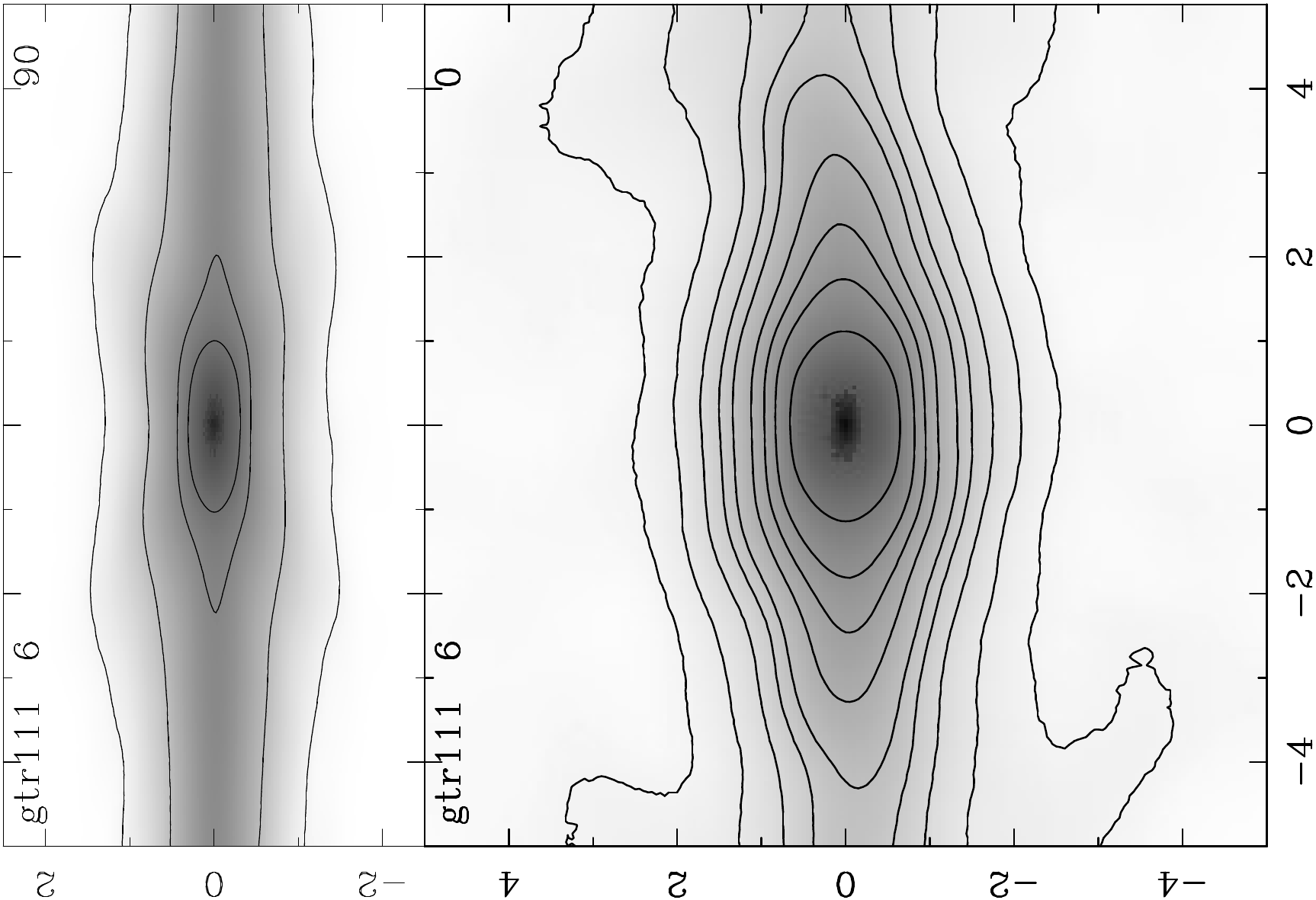}
  \caption{Grey-scale plots and isodensities of face-on (lower panels)
    and edge-on (upper panels) views of simulation gtr111. From left
    to right the corresponding times are 3.5, 4 and 6 Gyr. The
    grey-scale levels 
    are logarithmically spaced and the isodensities are at levels
    chosen to show best given morphological features discussed in the
    text. In the upper left corner 
    of each sub-panel we give the simulation name and snapshot time and in
    the upper right one the viewing inclination angle in degrees. Note
    that in the rightmost panel the linear scale is different from
    that of the other two panels, so as to follow
    the growth of the bar.     
}
\label{fig:early-times2}
\end{figure*}

We have argued that barlenses are simply the B/P/X component viewed
face-on. A corollary to this would be that there are no barlenses in
cases with no B/P/X. Is this indeed the case in our simulations?

To address this, we carefully examined our
simulations before the B/P/X formed. Let us first note that for
many simulations it 
is difficult to give precisely the time of B/P/X formation and that
errors of at least 0.5 Gyr and often considerably larger should be
expected. With that proviso let us look at a typical case in 
Fig.~\ref{fig:early-times2}. Simulation gtr111 has initially 50\% of
its baryons in the form of gas and the evolution of its bar strength and 
B/P/X strength with time are given in figs 1 and 2 of
\cite{Iannuzzi.Athanassoula.15}, respectively. In particular, these
figures show 
that the peanut starts growing shortly after $t$=4 Gyr, with a
best estimate of $t$=4.5 Gyr. Thus the right-hand panel of
Fig.~\ref{fig:early-times2}, at $t$=6 Gyr is well after the time when
the B/P/X started growing, as is asserted also by 
the corresponding upper subpanel, where the peanut structure is clear. At
this time the face-on view (lower subpanel of the right-hand column)
clearly displays a barlens component. At $t$=3.5 Gyr (left-hand column),
i.e. before the B/P/X started growing, 
this simulation shows no barlens component, while at $t$=4 (middle
panel) -- i.e. roughly at the moment where
the B/P/X starts growing, or slightly before it -- we see that the
barlens starts becoming visible. 

This and other examples from our simulations argue that, at least for
the models that we have 
analysed here, the barlens and the B/P/X grow roughly concurrently, but
shifts of the order of half or one Gyr can not be excluded. Early on,
at times when the B/P/X has clearly not started growing, no barlens is
visible in the simulations. This means that some of the observed bars,
the relatively youngest ones, will not have a barlens component, in good
agreement with the observation result of \cite{Laurikainen.SABBJ.13},
who found that not all barred galaxies have barlenses. Furthermore, it
argues that all
galaxies with a barlens component must be in their secular evolution
phase.

\subsection{The orbital structure of barlenses}
\label{subsec:orbits}

The first step to understanding the dynamics of any substructure is
to study the 
orbits that constitute it. No specific study of orbits in barlenses has
so far been made, but since these structures are the face-on view of the
B/P/X bulges, as we argued in this paper, the orbital
structure of the latter (discussed e.g. by \citealt{Pfenniger.84, 
Patsis.SA.02}, \citealt{Skokos.PA.02a} and
\citealt{Patsis.Katsanikas.14}; see also \citealt{Athanassoula.15} for
a review) should provide the necessary information. The most likely
building blocks for our case and 
the most widely discussed in the literature, will then be 3D families of
periodic orbits of the $x_1$ tree which bifurcate from the vertical
resonances of the 2D $x_1$ family and have stable parts of a
sufficiently large extent. The families that 
bifurcate at the lowest energies -- such as $x_1v_1$, $x_1v_3$, 
$x_1v_4$, or $x_1v_5$ -- are the
most likely candidates, since the outline of orbits from higher order
bifurcating families are less vertically extended and more extended 
along the bar major axis,
and thus less appropriate building blocks. An alternative would be the
$z3.1s$ family \citep{Skokos.PA.02b}, whose orbits have a morphology 
similar to that of the $x_1v_4$ family but is not related to
the $x_1$ tree. This, however, would be suitable
only in specific models, while B/P/X bulges are more generally found
in barred models. This alternative is therefore not very likely and we
will not consider it further here.

\begin{figure}
  \center
  \includegraphics[scale=0.3, angle=-90]{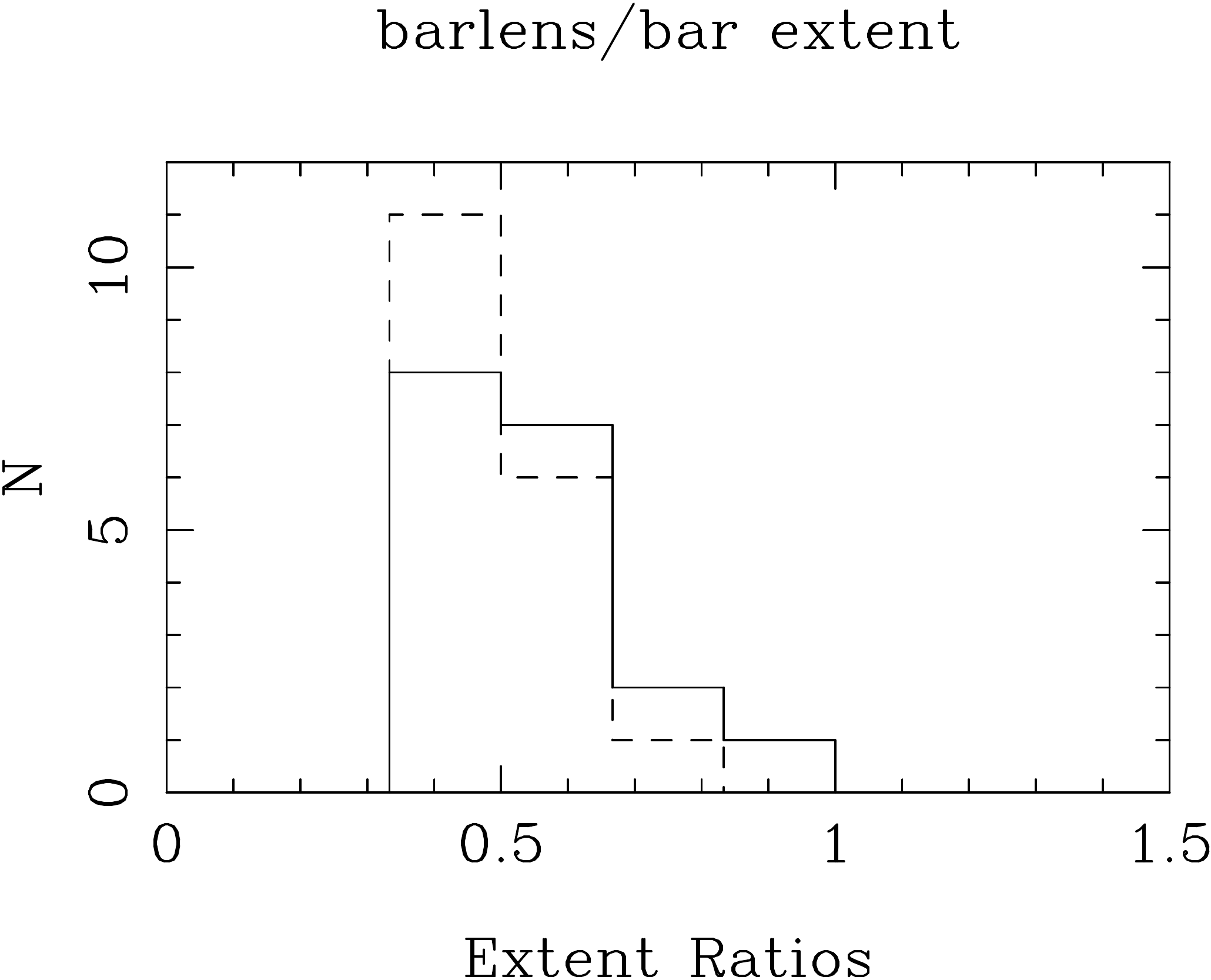}
  \caption{
Histogram of the ratio of the extent of the barlens to that of the
thin part of the bar. The solid line gives the histogram for the
extent ratios given in Table~\ref{tab:rellipsefits}, i.e. using the
same method as previously used for the NIRS0S observations, while the
dashed line gives the 
histogram of the same quantity, but with measurements obtained from
the face-on and side-on isodensity curves, as described in
Sect.~\ref{subsec:sideon}. 
    }
\label{fig:ratbl-histo}
\end{figure}

As mentioned above, orbits from families bifurcating from the $x_1$ at
low energies (low values of the Jacobi constant) are less extended along the
bar major axis compared to the thin part of the
bar than orbits from higher order families. Thus, by comparing the
horizontal extent of the barlens to that of the thin bar component,  
we should in principle be able to deduce which vertical resonance is
the main contributor to the barlens and where it is located. In
practice, however, this is not easy because the relative extent of the
orbits of various families is model dependent. In
Table~\ref{tab:rellipsefits} we gave values for the ratio of the
barlens extent to that of the thin bar component for the $t$=6 Gyr
snapshots of all simulations discussed here and in
Fig.~\ref{fig:ratbl-histo} we give a histogram of these values (solid
line). As already mentioned in Sect.~\ref{subsec:lengthrat-axrat}, 
all but one of the snapshots have extent ratios within the range
defined by the NIRS0S observations. The exception is gcs006, whose
measurement, as discussed in Sect.~\ref{subsec:ell-results}, is not
as reliable as the rest.  

Our next step will be to compare the values of the extent ratios with
the corresponding numbers found from orbital structure work. In
particular we want to compare them with the corresponding numbers in
tables 1 to 6 of \cite{Patsis.SA.02}. These set rough limits to the
extent ratios depending on the main family that constitutes the barlens. For
example for $x_1v_1$ the ratio of barlens to bar extent should be less
than 0.5, while for the other families it should be larger than 0.6 and
less than 1. As we will discuss below, there is always some uncertainty
in these numbers, because the orbits were calculated in a fiducial
potential and not in the potential of each given galaxy and, furthermore,
the measurements of the barlens and bar extent always have some
measuring error. 

Fig.~\ref{fig:ratbl-histo} shows that, making a 
conservative estimate  of a 10\% global uncertainty, only 3 out
of 18 models have L(bl) / L(bar) which is not compatible with the
$x_1v_1$ family. The three exceptions are gtr101, gtr117 and gcs006,
which have extent ratios of 0.77, 0.81 and 0.94, respectively. Note
that the measurements given for these three cases are less reliable
than the remaining ones, as discussed in Sect.~\ref{subsec:ell-results}
and \ref{subsubsec:decomp-results}. Thus, our results indicate that in
about 83\% of our simulations, and perhaps more, the $x_1v_1$ family
can be 
the backbone of the B/P/X/bl component. This number is not in good
agreement with the NIRS0S results, where this fraction is lowered to
about 54\%. The disagreement is not surprising. As already mentioned,
our simulations covered the gas fraction and halo shape parameter
space homogeneously, so as to better explain their effect on bar
formation and evolution. They did not attempt to model specific
galaxies, nor did they attempt to follow their distributions in
parameter space. Thus the 
distribution of any result, such as the extent ratios, can not be
expected to be similar in observations and in simulations. There is,
nevertheless, good agreement between the two because 
the ranges of values covered by observations and simulations are in
good agreement, as we already showed in Sect.~\ref{subsec:lengthrat-axrat}.     

In Sect.~\ref{subsec:sideon} we used the isodensity curves in the
side-on view to measure the B/P extent, and found that the results
agreed very well with those obtained for the barlens from the face-on view
given in Sect. \ref{sec:ellipsefits}. Similarly, the extent of the bar
can be measured from the isodensity curves in the face-on view. We
compared the latter 
results with the bar length measurements obtained with the method used in
the NIRS0S analysis (see Sect.~\ref{sec:ellipsefits}) and found
differences. These differences are not due to different personal
appreciation, but are systematic, due to differences between methods.  
Indeed the isodensity method places the end of the bar further out than
the ridge line of the ring. The median value of the ratio of the
bar length as obtained with the NIRS0S rule to that obtained from the
isodensity shapes is 0.86, and the mean 0.87, with an absolute
deviation of 0.11. We then calculated the ratios of barlens extent to
bar length using for both quantities the values obtained by the
isodensities and 
plot the result in Fig.~\ref{fig:ratbl-histo} (dashed line). As
expected, the values of the extent ratios are smaller, so the histogram is somewhat
shifted. Now only two simulations have a ratio beyond 0.6, namely
gtr101 and gcs006, with ratios 0.62 and 0.73, respectively. These
two simulations have no gas and their measurements are less accurate
than those of the rest. We may thus conclude that a very large fraction, 
if not all the simulations could have a B/P/X/bl feature compatible 
with the $x_1v_1$ family.  

Some care, however, has to be taken when using results from orbital
structure theory to make detailed quantitative comparisons with
observations. Bars have a very complex morphology and
geometry and, as a result, their potential has not been yet adequately 
modelled analytically. The most realistic of the existing
analytic models, the Ferrers models \citep{Ferrer.77}, have an
ellipsoidal face-on shape, which is not a good description of the real bar
shape \citep{Athanassoula.MWPPLB.90, Gadotti.09, Gadotti.11}, and,
viewed edge-on, do not have a B/P/X. Yet, for lack of any 
better potential/density model, 
Ferrers models have been used in most  orbital structure 
studies of the barlens/bar/bulge region \citep[e.g.][]{Athanassoula.BMP.83, 
Papayannopoulos.Petrou.83, Pfenniger.84, Athanassoula.92, Skokos.PA.02a, 
Skokos.PA.02b, Patsis.05}. The main qualitative results of
these studies should be applicable to real galaxies, but at
least some of their quantitative results may not be sufficiently
accurate for detailed comparisons with observations. It would
therefore be highly desirable to find analytic galactic potentials
that take into account both the 3D shape of the B/P/X/barlens
component and 
the rectangularity of the thin bar component, in order to make yet
more realistic orbital structure studies.    

\cite{Skokos.PA.02a} noted that the $(x,y)$ projections of the 3D
families of the $x_1$-tree, i.e. the $x_1v_n, n=1,2,3,..$ families, retain a
morphological similarity with the parent $x_1$ family at the same
energy, in particular in regions relatively near the 
bifurcating points. This could, at first sight, have been
considered to disagree with the fact that the face-on outlines of the
barlens are very different from that of the thin bar components, and
thus argue against 
the barlens picture that we propose here. It must, however, be kept in
mind that the orbital structure study of \cite{Skokos.PA.02a}, like
those of all other such studies, is carried out using Ferrers
potentials and does not include the potential of a separate barlens
components. This is a further argument underlining the necessity of
orbital structure studies using yet more realistic potentials.

\subsection{Discy bulges}
\label{discy}

We have so far discussed at length the links between B/P/X bulges
and barlenses and also mentioned tentatively that, if no appropriate
kinematic or photometric data are available, barlenses could be confused
with classical bulges. There is, however, yet another type of bulges
which we have not discussed yet,
namely discy pseudo-bulges \citep{Kormendy.Kennicutt.04,
  Athanassoula.05, Erwin.08}. These have the shape of a thin disc and
a large fraction of their mass is in gas and relatively young
stars. Could they, like the classical bulges, be mistaken for a
barlens? We think this is unlikely, for many reasons, of which
the most important is that 
their sizes are much smaller than that of the barlenses,
only of the order of 1 kpc or even smaller
\citep[e.g.][]{Kormendy.Kennicutt.04, Athanassoula.05}. Furthermore, their
populations are on average younger than that of barlenses.

Given their possible formation scenarios, it is expected that the
various types of bulges will very often co-exist
\citep{Athanassoula.05}  
and this was indeed shown to be true by observations
\citep{Kormendy.Barentine.10, Nowak.TESBD.10, Mendez.ADCA.14,
  Erwin.plus.15}. Could 
co-existing classical and discy bulges in galaxies with a bar but no
B/P/X bulge give structures having the
properties of barlenses? We have at our disposal no simulations with the adequate
structures -- i.e. no appropriate simulations with both classical and discy bulges --
to examine this question in depth. However, in view of the arguments we
gave above on the inadequacy of discy bulges to provide reasonable building
blocks for barlenses, this does not seem likely.    

Of course discy bulges and barlenses can co-exist (see references in the
previous paragraph). In such cases a number of structures of limited
radial extent, such as nuclear spirals, nuclear rings
and nuclear bars, will be seen in the central parts of the barlenses,
but will in fact be part of the discy bulge
(e.g. \citealt{Laurikainen.SABBJ.13}, L+14), not of the
barlens itself.

\section{Summary}
\label{sec:conclusions}
\indent

In this paper we used snapshots of barred disc galaxy simulations to
make images, which we then 
analysed using the same procedures and software as for the analysis of
real galaxy images in previous studies. We found that our 
simulations can produce components with properties comparable to
those of observed barlenses. By making a number of comparisons
between simulations and observations -- including morphology, radial
projected density profiles, shapes obtained from ellipse fits and
decomposition results -- we find very good agreement and thus reach the
conclusion that our simulations are sufficiently realistic to
describe accurately components such as barlenses.

Viewing our simulations both face-on and edge-on we were able to explore the nature of
the barlens component and to find that the {\it barlens and the B/P/X
bulge are one and the same component, but viewed from a different angle}.
We found this to be true not only for 
elongated barlenses, but also for the case of near-circular ones.

This result concerning the nature of the barlens component implies
that the distribution of the $z$-component of the 
velocity of stars in the barlens should correspond to that of a vertically 
thick object, while
that of the stars in the thin component should be 
that of a vertically thin one. It should be possible to
test this prediction by measuring the vertical velocity component 
along the bar major axis of appropriately chosen face-on barred
galaxies. One should then find a much higher
velocity dispersion $\sigma_z$ and a different structure of the higher
velocity moments in the thick barlens component compared to the
thin part of the bar \citep[][]{Iannuzzi.Athanassoula.15}. It is
not, however, necessary to expect that 
there should be a sharp transition between the two 
regions. Indeed, in Sect.~\ref{sec:global} we found that there is no
such sharp transition in the radial luminosity profile along the bar
major axis, and we attributed
this to the fact that the thin bar component continues inwards into
the barlens region. Thus, in the region surrounding the bar major
axis and within the barlens component there will be contributions
from both components, leading to a smooth transition.

Our result has a further implication, namely that it can set a
rough lower limit of the bar age and a very clear one on the phase of
evolution the bar is in. Indeed, the fact that the barlens and the
boxy/peanut bulge are one and the same component implies that for the
barlens to exist, the B/P/X must 
have already formed. Therefore, the bar must be in the secular evolution phase,
i.e. between one and a few Gyr must have elapsed from the moment it
started forming, as was discussed in Sect.~\ref{formation}. 

For all the simulations discussed here, the most realistic barlenses 
were found when there is between 20 and 80 per cent gas in the initial
conditions. Collisionless simulations, i.e. simulations whose initial
conditions contain no gas, have barlenses which are too extended
compared to observations and
it is not even clear whether using a second bar component in the
decomposition improves the fit. This could be expected because 
real galaxies will normally contain some gas when the bar grows,
except for the rather unlikely case when all the gas is consumed
before the bar starts growing and there is no accretion to add more
(AMR13, \citealt{Athanassoula.14}).  
At the other extreme, simulations with
initially 100\% gas have very short thin bar components, so that they
are more reminiscent of barlenses in non-barred galaxies. Note also
that, for such high initial gas fractions, the trend between the
fraction of the total flux which is in the 
barlens component and the bar strength has a smaller slope than the
observations and is considerably offset towards larger barlens fluxes.  

The morphology of the barlens component is such 
that in some cases it can be mistaken for a classical
bulge, unless high resolution and high quality photometry and/or
kinematics are available. Furthermore, not including a
barlens in decompositions can artificially increase the
bulge-to-total ratio by, on average, a factor of 3.5 (L+14). This is
due to the classical bulge compensating for the lack of the barlens
component. 
Thus the
number of galaxies with no classical bulge is higher than
what has been so far assumed and also in many galaxies the
contribution of the classical bulge to the total mass budget is
substantially lower than what has been so far estimated. This can
introduce errors in dynamical 
studies of some individual galaxies and a bias in 
a number of statistical studies. We propose that photometry and
particularly kinematics can be used to distinguish
between classical bulges and barlenses and thus avoid misclassifications. 

We also give a rule of thumb to obtain the extent of the barlens
component in the direction of the bar major axis, for relatively low
inclination galaxies. This, however, will 
not work for all possible morphologies, as for example cases where
the isodensities of the bar are concentric thick ellipses. We
showed that in most of the simulations considered in this paper,
the B/P/X/bl component is compatible with a $x_1v_1$ backbone,
provided the Ferrers models can be considered sufficiently
realistic. On the other hand, the fraction of real
galaxies compatible with an $x_1v_1$ barlens backbone could be
considerably smaller, of the order of 
50--60\%.

We also found a trend between the fraction of the total flux that is in
the barlens component and the strength of the
bar, which argues that
during the secular evolution phase, as the bar grows stronger the
barlens component also becomes more important, so that it represents
an increasing fraction of the baryonic mass. This trend is
confirmed by observations. The quantitative comparison is best for a
model with 50\%, rather than 100\% initial gas fraction, which is
indeed in better agreement with observations 
\citep[e.g.][and references therein]{Genzel.p.14}. 

All our tests confirm that barlenses can be considered as face-on
  B/P/X bulges. We do not, however, claim that this is the only possible
alternative, i.e. that there is no other way to explain barlenses. 
Indeed, proving uniqueness is very difficult, if not impossible for
the formation of galaxies, or of their components and we can not
exclude 
that another alternative will be brought up in the future. We did,
nevertheless, outline some of the serious problems that alternatives
based on a discy bulge component would face.

Our work also underlines the necessity of a further improvement. 
Namely, a considerable part of orbital structure
studies and of gas flow calculations should be revisited using yet
more realistic 
potentials, which take into account the complex 3D structure of bars
and include the vertical structure of the B/P/X/bl component.

\section*{Acknowledgements}

We thank the referee for his/her comments, Volker Springel for
making available to us 
the version of \textsc{gadget} used here and Jean-Charles Lambert for
the \textsc{GLNEMO} software (http://projets.oamp.fr/projects/glnemo2). 
All authors acknowledge financial support to the DAGAL network from the People
Programme  (Marie Curie Actions) of the European Union's Seventh
Framework Programme FP7/2007-2013/ under REA grant agreement number
PITN-GA-2011-289313. EA and AB also acknowledge financial support from the
CNES (Centre National d'Etudes Spatiales - France) and from the
`Programme National de Cosmologie and Galaxies' (PNCG) of CNRS/INSU,
France. EA acknowledges the use of HPC resources from GENCI-
TGCC/CINES (Grants 2013 - x2013047098 and 2014 - x2014047098). EL and
HS acknowledge support from the Academy of Finland. This
research is based in part on observations made with the Spitzer Space
Telescope,  
which is operated by the Jet Propulsion Laboratory, California Institute of
Technology under a contract with NASA. 

Funding for the SDSS and SDSS-II has been provided by the Alfred P. Sloan Foundation, the Participating Institutions, the National Science Foundation, the U.S. Department of Energy, the National Aeronautics and Space Administration, the Japanese Monbukagakusho, the Max Planck Society, and the Higher Education Funding Council for England. The SDSS Web Site is http://www.sdss.org/.
The SDSS is managed by the Astrophysical Research Consortium for the Participating Institutions. The Participating Institutions are the American Museum of Natural History, Astrophysical Institute Potsdam, University of Basel, University of Cambridge, Case Western Reserve University, University of Chicago, Drexel University, Fermilab, the Institute for Advanced Study, the Japan Participation Group, Johns Hopkins University, the Joint Institute for Nuclear Astrophysics, the Kavli Institute for Particle Astrophysics and Cosmology, the Korean Scientist Group, the Chinese Academy of Sciences (LAMOST), Los Alamos National Laboratory, the Max-Planck-Institute for Astronomy (MPIA), the Max-Planck-Institute for Astrophysics (MPA), New Mexico State University, Ohio State University, University of Pittsburgh, University of Portsmouth, Princeton University, the United States Naval Observatory, and the University of Washington.
 
\bibliographystyle{mn2e.bst}
\bibliography{barlens1_subm4_comments.bbl}

\label{lastpage}

\end{document}